\documentclass[letterpaper]{mn2e}
\usepackage{graphicx,times}

\def \sanch{S\'anchez-Bl\'azquez}

\def \kms {${\rm{km}\,\rm{s}^{-1}}$}

\def \ha   {H$\alpha$}
\def \hb   {H$\beta$}

\def \hd   {H$\delta$}

\def \caii   {Ca\,{\sc ii}}

\def \afe  {[$\alpha$/Fe]}
\def \nafe  {$\alpha$/Fe}

\def \zh  {[Z/H]}

\def \apjs{ApJS}

\def \apj{ApJ}
\def \aj{AJ}
\def \mnras{MNRAS}

\def \aanda{A\&A}

\title[Driving parameters of stellar populations]
{Ages and metallicities for quiescent galaxies in the Shapley Supercluster: Driving parameters of the stellar populations}
\author[Russell J. Smith et al. ]
{Russell J. Smith$^{1}$, John R. Lucey$^{1}$, Michael J. Hudson$^{2}$\\
$^1$Department of Physics, University of Durham, Durham DH1 3LE, United Kingdom\\
$^2$Department of Physics and Astronomy, University of Waterloo, Waterloo, Ontario N2L 3G1, Canada}

\pagerange{\pageref{firstpage}$-$\pageref{lastpage}} \pubyear{2009}

\begin{document}

\label{firstpage}

\maketitle

\begin{abstract}
We use high signal-to-noise spectroscopy for a sample of 232 quiescent galaxies in the Shapley 
Supercluster, to investigate how their stellar populations depend on velocity dispersion ($\sigma$), luminosity and stellar mass.
The sample spans a large range in velocity dispersion (30--300\,\kms) and in luminosity ($M_R$ from --18.7 to --23.2).  
Estimates of age, total metallicity (Z/H) and $\alpha$-element abundance ratio (\nafe) were derived from absorption-line analysis,
using single-burst models of Thomas and collaborators.  
Using the Rose \caii\ index, we conclude that recent star-formation (``frosting'') events are not responsible for the intermediate
ages observed in some of the galaxies. 
Age, Z/H and \nafe\ are correlated positively with 
velocity dispersion, but we also find significant residual trends with luminosity: at given $\sigma$, the brighter galaxies
are younger, less $\alpha$-enriched, and have higher Z/H. 
At face value, these results might suggest that the stellar populations depend on stellar mass as well as on velocity dispersion. 
However, we show that the observed trends can be reproduced by models in which the stellar populations depend systematically 
{\it only on} $\sigma$, and are independent of stellar mass $M_*$.
For age, the observed luminosity correlation arises because young galaxies are brighter, at fixed $M_*$. 
For metallicity, the observed luminosity dependence arises because metal-rich galaxies, at fixed mass, tend also to be younger, and hence brighter. 
We find a good match to the observed luminosity correlations with Age\,$\propto\sigma^{+0.40}$, Z/H\,$\propto\sigma^{+0.35}$, $\alpha/{\rm Fe}\,\propto\sigma^{+0.20}$, 
where the slopes are close to those found when fitting traditional scaling relations.
We conclude that the star formation and enrichment histories of galaxies are determined primarily by the depth of their gravitational
potential wells. The observed residual correlations with luminosity do not imply a corresponding dependence on stellar mass. 
\end{abstract}

\begin{keywords}
galaxies: elliptical and lenticular, cD ---
galaxies: evolution 
\end{keywords}

\section{Introduction}

Numerous studies have explored how the stellar content of passive galaxies\footnote{
By the terms passive/quiescent galaxies, we loosely include many partially-overlapping definitions, i.e. galaxies which have ceased forming stars, 
lie in the red envelope of the optical colour distribution, have little or no line emission in their spectra, have early-type morphology, etc.}
depends systematically on ``mass''. Along the passive sequence, more massive galaxies are observed to be redder, and have stronger metal absorption and weaker
hydrogen features in their spectra (e.g. Faber 1973; Bower, Lucey \& Ellis 1992). Physical explanations for these trends have been sought 
using increasingly sophisticated stellar population synthesis models (e.g. Worthey 1994; Thomas, Maraston \& Bender 2003; Schiavon 2007), 
and ever improving observational datasets (e.g. Trager et al. 2000; Kuntschner et al. 2001; Caldwell et al. 2003; Nelan et al. 2005; 
Thomas et al. 2005; Bernardi et al. 2006; Graves et al. 2007). These studies have converged on a scenario in which more massive passive galaxies 
on average harbour older stars, with higher total metallicity, and with higher abundance of $\alpha$ elements relative
to Fe, as compared to less massive galaxies. 

While the strong mass-dependence of passive galaxy properties is widely appreciated, it is less clear which aspect of 
``mass'' is really dominant in establishing the observed correlations. From simple physical arguments we might expect metallicity to 
depend on the retention of gas within the galaxy halo during a galactic wind phase (e.g. Arimoto \& Yoshii 1987), and to be primarily correlated with 
escape velocity, and thus with velocity dispersion $\sigma$. Alternatively, if the amount of metals and stars produced both
reflect the overall efficiency of star-formation, then metallicity might instead be correlated primarily with stellar mass, 
and hence with luminosity. Because luminosity and velocity dispersion are mutually correlated through the Faber--Jackson (1976, FJ) relation, 
and because stellar mass is inferred rather than measured directly, it is not trivial to distinguish the relative importance of these 
mass proxies in driving the relations. By exploiting the substantial scatter in the FJ relation, a number of previous works have indicated 
that luminosity and velocity dispersion are not interchangeable as predictors of the
stellar populations (e.g. Bernardi et al. 2005; Gallazzi et al. 2006). 
In particular, Graves, Faber \& Schiavon (2009) recently analysed a sample of galaxies from the 
Sloan Digital Sky Survey (SDSS), concluding that important correlations with stellar mass persist after accounting for the trends with 
velocity dispersion. 

In this paper, we address these issues using results from high signal-to-noise spectroscopy for galaxies in the Shapley Supercluster. 
The observations were described in detail by
Smith, Lucey \& Hudson (2007, hereafter Paper I), which also presented measurements of 
absorption line strengths for a sample of quiescent supercluster member galaxies. Here we employ
the data to determine ages, total metallicities (Z/H) and $\alpha$-element abundance ratios (\nafe) 
for each galaxy in the sample, and analyse their dependence on luminosity and velocity dispersion.

Section~\ref{sec:agemet} provides a brief review of the sample definition and spectroscopic data, 
and describes the estimation of single-burst equivalent stellar population parameters. We also provide here
a test for consistency with the broadband colours, discuss the evidence against widespread
recent secondary bursts of star-formation, and describe the corrections for metallicity gradients. 
In Section~\ref{sec:agemass}, we present the correlations
of age, Z/H and \nafe\ with velocity dispersion and luminosity. In Section~\ref{sec:stelmass}, motivated by the significant correlations of
the stellar population parameters with luminosity, at fixed velocity dispersion, we investigate whether these require an 
underlying correlation with stellar mass. Section~\ref{sec:discuss} compares the results to previous findings, especially
focusing on the work of Graves et al. (2009), and our main conclusions are summarized in Section~\ref{sec:concs}. 

The stellar population parameters as determined here have been used by Rawle et al. (2008a) to study the dependence of ultra-violet colours
on the stellar populations, and by Gargiulo et al. (2009) to analyse stellar population effects in the Fundamental Plane
residuals. An alternative analysis of a subset of the Paper I data, using the method of Graves \& Schiavon (2008) has been presented by
Smith et al. (2009a). A more thorough exploration of complex star-formation history models is provided by Allanson et al. (2009). 
A later paper will treat the correlations between stellar population parameters, e.g. the age--metallicity--mass relation. 

Throughout this paper, we adopt cosmological parameters $(\Omega_M,\Omega_\Lambda,h)=(0.3,0.7,0.7)$.
For reference, at the redshift of Shapley ($z=0.048$), one arcsecond corresponds to 0.97\,kpc, 
and the distance modulus is $m-M=36.65$. 

\section{Stellar population parameters}\label{sec:agemet}

\subsection{Data}\label{sec:data}

Paper I provides a full description of the observations and data processing. Here, we provide only a summary of 
the main points. 

The initial galaxy sample was drawn from photometric catalogues of the NOAO Fundamental Plane Survey 
(NFPS, Smith et al 2004), covering the central 40$\times$40 arcmin$^2$ region in each of three clusters, 
Abell 3556, Abell 3558 and Abell 3562, in the core of the Shapley Supercluster. 
Spectra were obtained using 
the fibre-fed dual-beam AAOmega spectrograph at the Anglo-Australian Telescope. The fibres sample an aperture of 2\,arcsec diameter. 
Within the region covered by the NFPS catalogue, 
we observed some 60\,per cent of all $R<18$ galaxies, with little dependence on magnitude. In the blue arm of the spectrograph, 
the spectra cover the main Lick indices, up to Fe5406 at the red end, with a spectral resolution of 3.2\,\AA\ FWHM. 
The red arm was used to record the \ha\ region, at 1.9\,\AA\ resolution, for the purpose of detecting nebular emission. 
The total integration time was $\sim$8\,hours per galaxy, resulting in high signal-to-noise ratio 
(median S/N\,$\approx$\,60\,\AA$^{-1}$ at 4400--5400\,\AA). 

The primary parameters measured from the spectra are the redshift, $cz$, the velocity dispersion, $\sigma$, equivalent
widths of various emission lines after removing the stellar continuum, and the Lick absorption line indices. 
For some galaxies, the velocity dispersion is ``unresolved'', i.e. indistinguishable from zero in our spectra. 
The absorption line indices were transformed to the spectral resolution of the Lick system and corrected for the effects of
velocity broadening. 
A final sample for stellar population analysis was defined which excludes galaxies with \ha\ emission equivalent width above 
0.5\,\AA\ (which indicates likely contamination of the Lick Balmer-line indices), and those which lie outside of the
adopted redshift range for the supercluster ($cz=11670-17233$\,\kms). The final sample comprises 232 galaxies, of which
198 have measured velocity dispersion and 34 are unresolved. 
The analysis presented in this paper is based on the line-strengths and supporting data provided in 
tables 2--4 of Paper I.

Note that the sample was not selected explicitly according to morphology at any stage, and is therefore
representative of the passive population overall. 
Although no selection on colour was applied, our \ha\ selection very
effectively restricts the sample to the red sequence in the $B-R$ colour magnitude relation (see figure~1 of Paper I).

\subsection{Linestrength diagrams}\label{sec:gridplots}

\begin{figure*}
\includegraphics[angle=270,width=180mm]{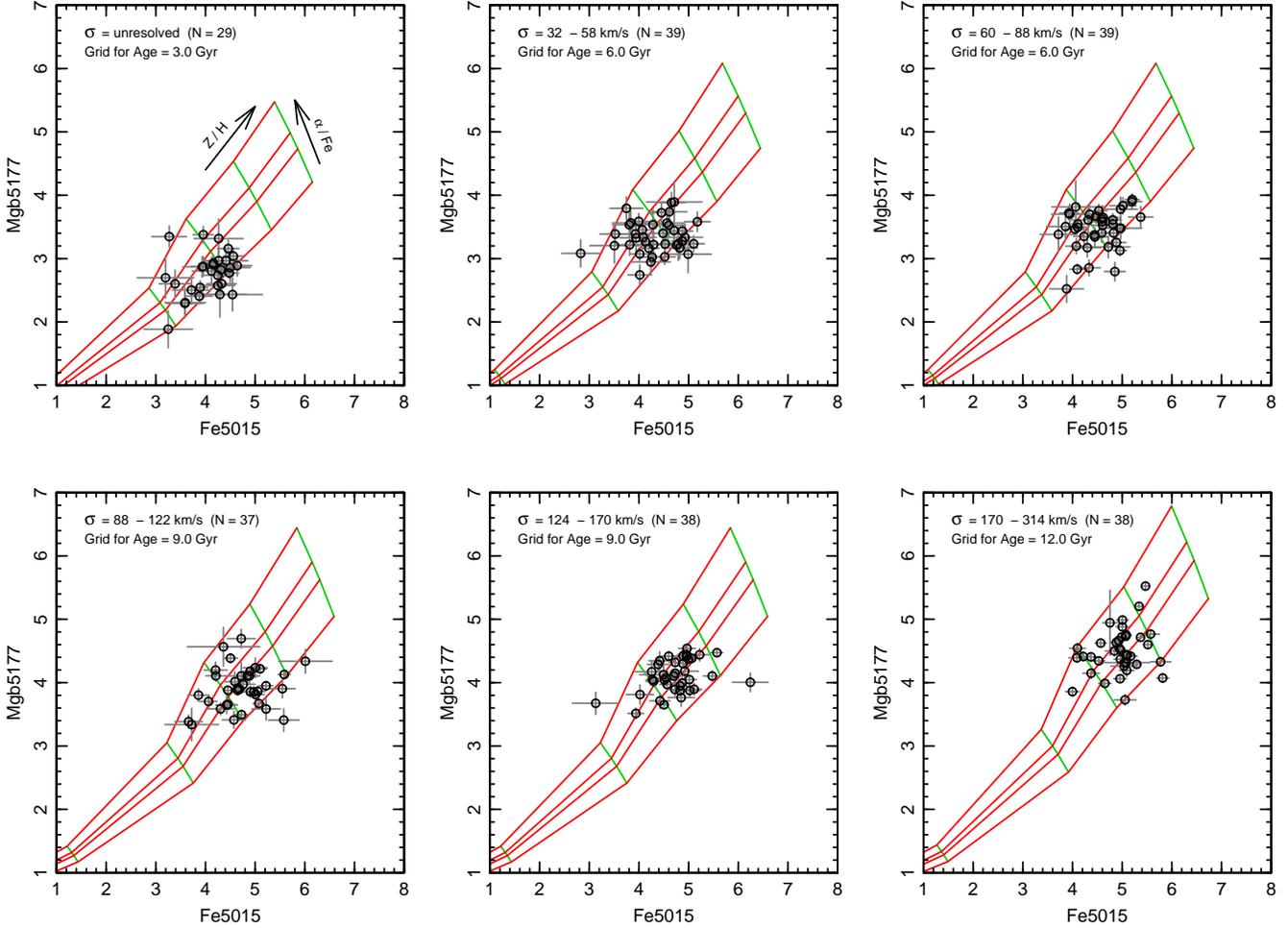}
\caption{Index--index planes showing the \nafe-sensitive index pair Fe5015 and Mgb5177. 
Each panel shows galaxies within a given interval in velocity dispersion. 
For clarity, we omit the $\sim$4\,per cent of galaxies having $S/N<20$, and 
two other galaxies with large error bars in Fe5015 due to incomplete data.
The model grids are from Thomas et al. 
(2003, 2004), with zero-points re-calibrated from the most massive galaxies (see Section~\ref{sec:thominvert}). The grids show lines of 
constant \afe\ = $0.0, +0.2, +0.3, +0.5$ (red) and of constant \zh\ = $-0.33, 0.00, +0.35, +0.67$ (green).
For this comparison, 
the grids are drawn for a fiducial age in each panel, as indicated in the legend. The fiducial ages are estimated from visual
examination of Figure~\ref{fig:agegrid}.}
\label{fig:afegrid}
\end{figure*}

\begin{figure*}
\includegraphics[angle=270,width=180mm]{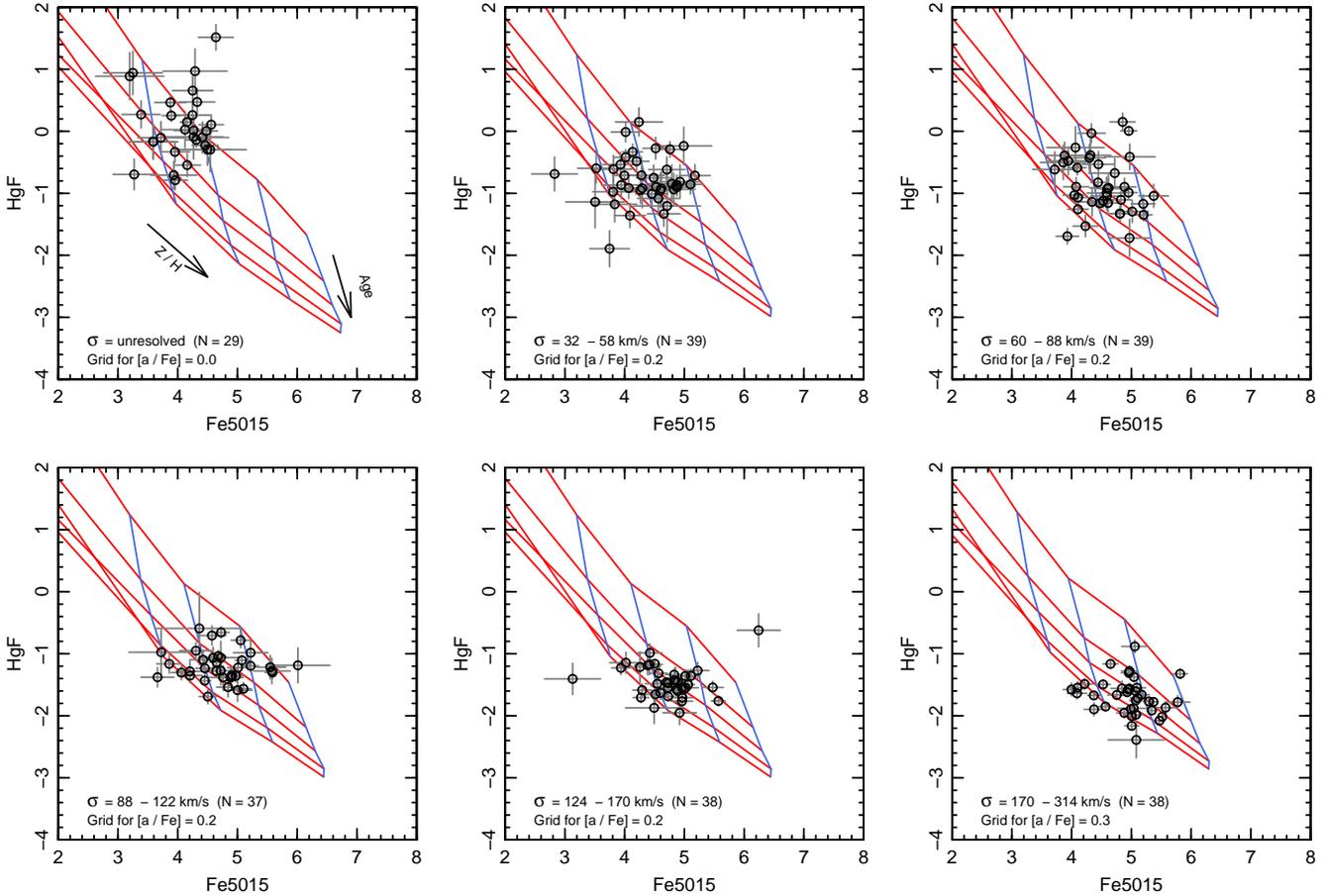}
\caption{Index--index planes showing the age-sensitive index pair Fe5015 and HgF.
Each panel shows galaxies within a given interval in velocity dispersion.
For clarity, we show omit the $\sim$4\,per cent of galaxies having $S/N<20$. 
The model grids are from TMBK, with zero-points re-calibrated from the most massive galaxies (see Section~\ref{sec:thominvert}). The grids show 
lines of constant \zh\ = $-0.33, 0.00, +0.35, +0.67$ (blue)
and of constant age = $3, 6, 9, 12, 15$\, Gyr (red). 
For this comparison, 
the grids are drawn for a fiducial \nafe\ for each bin, as indicated in the legend. The fiducial \nafe\ is estimated from visual
examination of Figure~\ref{fig:afegrid}.}
\label{fig:agegrid}
\end{figure*}

\begin{figure*}
\includegraphics[angle=270,width=180mm]{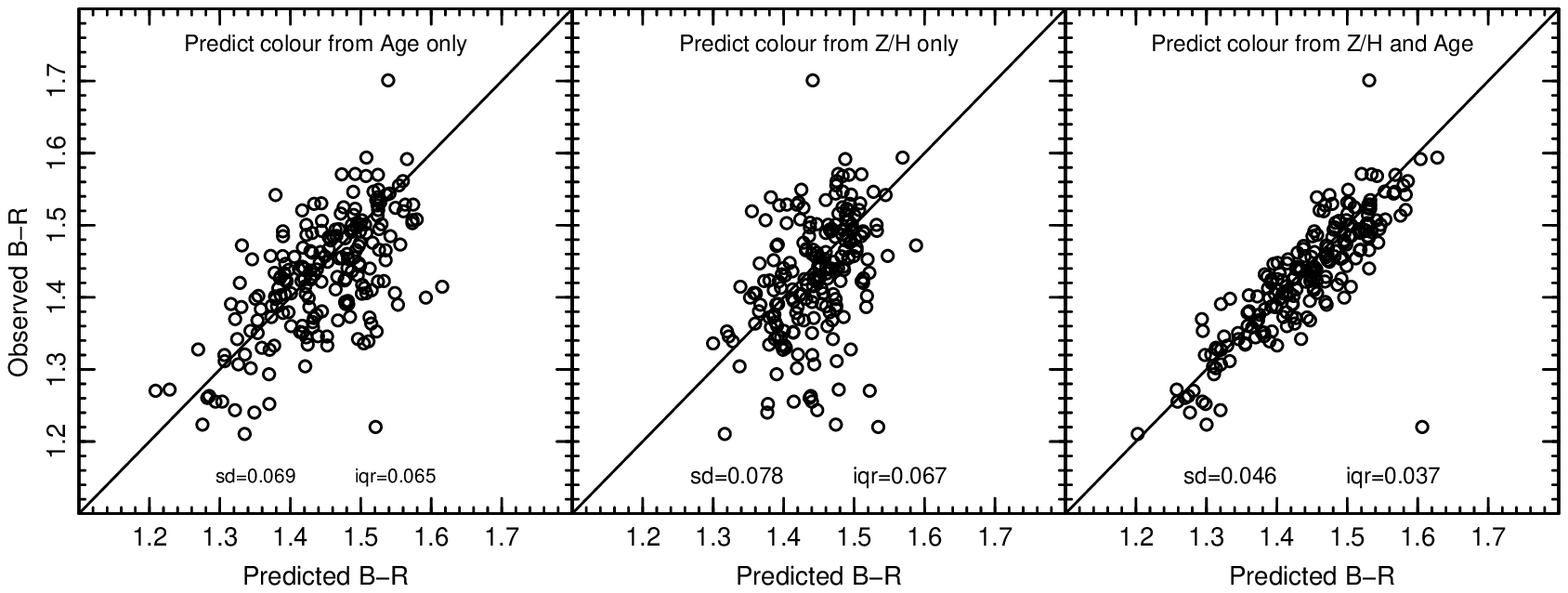}
\caption{Comparison of colours predicted from the spectroscopic parameters against the observed colours 
from the SOS catalogue of Mercurio et al. (2006). The observed colours, which are identical in all panels,
are within an aperture of 4.4\,arcsec diameter, i.e. larger than the 2\,arcsec diameter spectroscopic aperture. 
In the left panel, the colours are predicted via the 
SSP models of Maraston (2005), using the measured age, but assuming a common metallicity of \zh=0.11 (the median value
over the sample). In the centre panel, the measured ages are used, but a common age (6.5\,Gyr) assumed. The right panel uses
the measured values for both age and metallicity to predict the colours. 
The solid line shows equality; the spread is is quantified by the standard deviation (sd) and 
more robustly by the 68 per cent interval (iqr).}
\label{fig:modelsos}
\end{figure*}

This and the following section describe the process of transforming the set of index measurements for each galaxy into a set of physical parameters, 
by comparison with stellar population models. This ``grid-inversion'' approach has of course been applied by many 
other studies in the field, including some fairly sophisticated implementations (e.g. Proctor, Forbes \& Beasley 2004; Kelson et al. 2006; 
Graves \& Schiavon 2008). 
Here, the absorption line data will be interpreted using simple stellar population (SSP) models by Thomas and collaborators (Thomas, Maraston \& Bender 2003; 
Thomas, Maraston \& Korn 2004, hereafter collectively TMBK). Constraints on non-SSP formation histories will be discussed in Section~\ref{sec:compo}.

For the TMBK models, just three non-redundant index measurements are sufficient to invert the
model grid and recover estimates of log(Age), \zh\ and \afe. 
If more indices are available, a $\chi^2$ minimization provides an estimate of the SSP which best reproduces the observations. 
Such redundancy may be desirable both to suppress random errors, and to average over systematic error sources 
(e.g. the dependence of indices on chemical abundances beyond those allowed to vary in the models).
For age-sensitivity, at least one index should measure one of the Balmer lines. 
In principle, any two metal lines with different $\alpha$-sensitivity could be used to distinguish 
Z/H from \nafe\ effects. 
With the advent of improved models which follow the light-element abundances separately (e.g. Schiavon 2007), 
it is important to be clear which elements are contributing to the measured \nafe. 
Mgb5177 is the only index employed as an $\alpha$-abundance indicator, and although we will use the notation
\afe\ in the context of the TMBK models, our measurements primarily reflect $[$Mg/Fe$]$. 

Our choice of Fe-sensitive index is limited because the widely-used Fe5270 and Fe5335 features are contaminated by
5577\,\AA\ night-sky emission line in most of our spectra. The other red Fe index, Fe5406, may
be unreliable because it is close to the spectrograph dichroic cut-off. Among the bluer lines, Fe4668 is also known as C$_2$4668 because it
is dominated by the Swan C$_2$ band. Since the abundance behaviour of C may be distinct from both 
Fe and Mg (e.g. \sanch\ et al. 2006a; Clemens et al. 2006; Smith et al. 2009b), we prefer not to use Fe4668 here. 
Among the remaining indices, Fe4383 and Fe5015 are the most promising, with good sensitivity to Fe but not to 
the light elements (see Tripicco \& Bell 1995). 

Prior to describing the galaxy-by-galaxy grid inversion results, we present a visual impression of the 
data in the space defined by the Mgb5177, Fe5015 and HgF indices. 
Figures~\ref{fig:afegrid} and \ref{fig:agegrid} show projections of the model grid and the 
data, divided into six bins according to the velocity dispersion. 
The first bin contains the galaxies with unresolved velocity dispersion, while the remaining bins represent roughly
equal fractions of the galaxies with measured $\sigma$. 
In Figure~\ref{fig:afegrid}, the Mgb5177 vs Fe5015
plane is displayed; the location of a galaxy in this diagram reflects primarily its metallicity and \nafe, but
also to a lesser extent its age. The over-plotted grids are extracted from the TMBK models, 
with age chosen as appropriate for each $\sigma$ bin (determined by consideration of Figure~\ref{fig:agegrid}). 
From Figure~\ref{fig:afegrid}, the systematic increase of both Z/H and \nafe\ with increasing 
$\sigma$ is apparent, although there is substantial intrinsic scatter within each bin. In particular, in each bin there are galaxies
spanning the full range 0.0--0.5 in \afe\ covered by the models. 
The age effects are evident in Figure~\ref{fig:agegrid}, which shows HgF versus Fe5015. 
Again, the models are shown in two-dimensional projection with the third variable, in this case \afe, chosen
to match the average value inferred from Figure~\ref{fig:afegrid}. The systematic trend of Z/H is visible as the tendency
for the cloud of points to shift along the long axis of the grid with increasing velocity dispersion. A trend of 
increasing age is also seen in the averages. Within each bin there is substantial scatter, with galaxies
spanning the age range 3--15\,Gyr, clearly in excess of the measurement errors. 
In some of the panels it can be seen that the galaxies which are younger than average also have higher metallicity than average, 
i.e. there is an anti-correlation of age and metallicity at fixed velocity dispersion (Trager et al. 2000).

\subsection{Grid inversion}\label{sec:thominvert}

\begin{figure}
\includegraphics[angle=270,width=85mm]{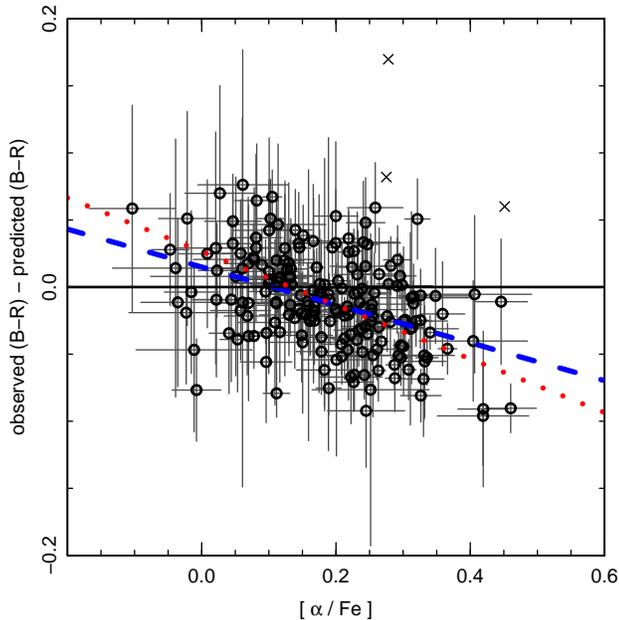}
\caption{Residual of the observed colour, relative to the colour predicted from the spectroscopic age and total metallicity, 
compared to the spectroscopic estimate of \afe. The dashed (blue) line shows a fit to the data after iterative rejection of outliers (crosses, without error bars). 
The most discrepant outlier NFPJ132802.6--314521 lies beyond the plot limits. 
The dotted (red) line of slope $-0.2$ represents
predictions from Coelho et al. (2007) for the effect of $\alpha$ enhancement on SSP colour.
The vertical error bars are dominated by contributions from the (anti-correlated) age and 
metallicity errors in the predicted colours. 
}\label{fig:afemodelsos}
\end{figure}

The formal grid inversions are performed using a non-linear $\chi^2$ minimisation over the six indices 
Mgb5177, Fe5015, Fe4383, Hbeta, HgF, HdF. 
The predicted values for each index are represented by a cubic-spline surface in the log(Age)--\zh--\afe\ space, 
passing through all grid points and interpolating 
between. Using a spline interpolation ensures continuous derivatives at cell boundaries; the
spline is constrained to reduce to linear extrapolation beyond the grid limits. 
Because the model grids are tilted with respect to the axes defined by the indices, the errors in recovered 
age and metallicity are correlated. We extract the full error covariance matrix from the $\chi^2$ surface, in order 
to propagate the correlated errors. 

Recalling that the indices were measured on (approximately) flux-calibrated spectra, and that no attempt was made to 
match the Lick system response curve, we expect some small systematic offsets between the models and the observed indices. 
Because the galaxies are all at a common distance, the indices are always measured at approximately the same observed wavelength, 
and any offsets due to different flux calibrations should thus be similar for all galaxies. 
If uncorrected, the offsets would generate zero-point shifts in the relations we aim to study, e.g. the Z/H-vs-$\sigma$ relation. 
A second-order effect is that, because the model grids are not perfectly parallel and linear, especially over
large ranges in age and metallicity, the slopes of the correlations could be affected by index offsets. 

A common approach is to observe stars from the original Lick stellar library to establish corrections to 
the measured indices (usually limited to a constant zero-point shift). This is observationally expensive, and
overlooks the possibility of zero-point uncertainties in the model predictions, which may be comparable to the 
observational offsets. Instead, we adopt a practical strategy of calibrating such that the most massive galaxies,
on average, yield stellar population parameters that are similar to those obtained in previous studies. 
Specifically we apply a rigid shift to the model grid, such that the median index-set for the 24 galaxies with $\sigma>200$\,\kms\ 
corresponds to the prediction for (Age, \zh, \afe) = (10.8\,Gyr, 0.24, 0.28). These values are taken from the high-$\sigma$ bin of 
Nelan et al. (2005), but they are characteristic of most other studies for giant ellipticals, including carefully Lick-calibrated work.
The index shifts, applied to the models (not the observations),  are:
(Fe5015:~--0.69),
(Hbeta:~--0.10),
(HdF:~--0.01),
(HgF:~--0.05),
(Mgb5177:~--0.10),
(Fe4383:~--0.22), 
in the sense that the re-calibrated indices are in all cases smaller than those predicted by the published models. 
We emphasize that only the average high-mass galaxy properties have been fixed in this way. Relative differences
in their populations are preserved, so that measurements of the age scatter at high mass, for instance, are meaningful.
Moreover, the relative properties of high- and low-mass galaxies are of course maintained, so that the slopes of the
stellar population scaling relations are unchanged, to first order\footnote{
As a robustness test, we have also estimated the stellar population parameters after applying
equivalent calibrations based on the Nelan at al. averages in bins at lower masses. Although this introduces overall shifts in the derived parameters
(e.g. 5--10\,per cent in the age zero-point), we find that the effect of this change in calibration is not systematically correlated with
luminosity or velocity dispersion. This test confirms that the coefficients of the scaling relations derived in Section~\ref{sec:agemass}
are stable (to within $\la$0.01) with respect to the details of the calibration scheme.}.
The calibration scheme involves only six degrees of freedom, compared 
to $\sim200\times6$ independent data.

In Table~\ref{tab:agefits} we present the derived stellar population parameters from the six-index inversion, as used in 
analysing the population trends in the following sections. 
Table~\ref{tab:agefits} includes the fit results for all 232
galaxies from Paper I. For the analyses in the following sections, we exclude ten of these galaxies which have low signal-to-noise spectra
($S/N<20$ per \AA\ at rest-frame 4400--5400\,\AA). Velocity dispersion measurements are required for the analysis in Section~\ref{sec:agemass}; removing objects with
unresolved $\sigma$ leaves 193 galaxies for use in the fits. 
The typical formal errors are 14\,per cent in age, 0.05\,dex in Z/H and 0.04\,dex in $\alpha$/Fe, for galaxies with $\sigma=50-100\,$\kms, 
and half of these values for galaxies with $\sigma>150\,$\kms.
As a function of signal-to-noise ratio, the errors are well represented by 
$\varepsilon_{\rm [Z/H]}=2.5 (S/N)^{-1}$ , 
$\varepsilon_{\rm log(Age)}=2.7 (S/N)^{-1}$ , and 
$\varepsilon_{\rm [\alpha/Fe]}=2.0 (S/N)^{-1}$. 

To assess the systematic uncertainties associated with our choice of indices, we have investigated how the results vary when different sets
of three indices are used to invert the model grids. In all cases the index set includes Mgb5177, one Fe-sensitive index (Fe5015 or Fe4383) and one
Balmer index (Hbeta, HgF or HdF), giving six possible triplets. Relative to our default six-index inversion, these triplets yield zero points differing by
$\la$0.10\,dex in age, $\la$0.05\,dex in Z/H, and $\la$0.03\,dex in $\alpha$/Fe. The differences between the results do not depend significantly on $\sigma$ 
or luminosity, hence the slopes of the scaling relations are quite robust.

\begin{figure}
\includegraphics[angle=270,width=80mm]{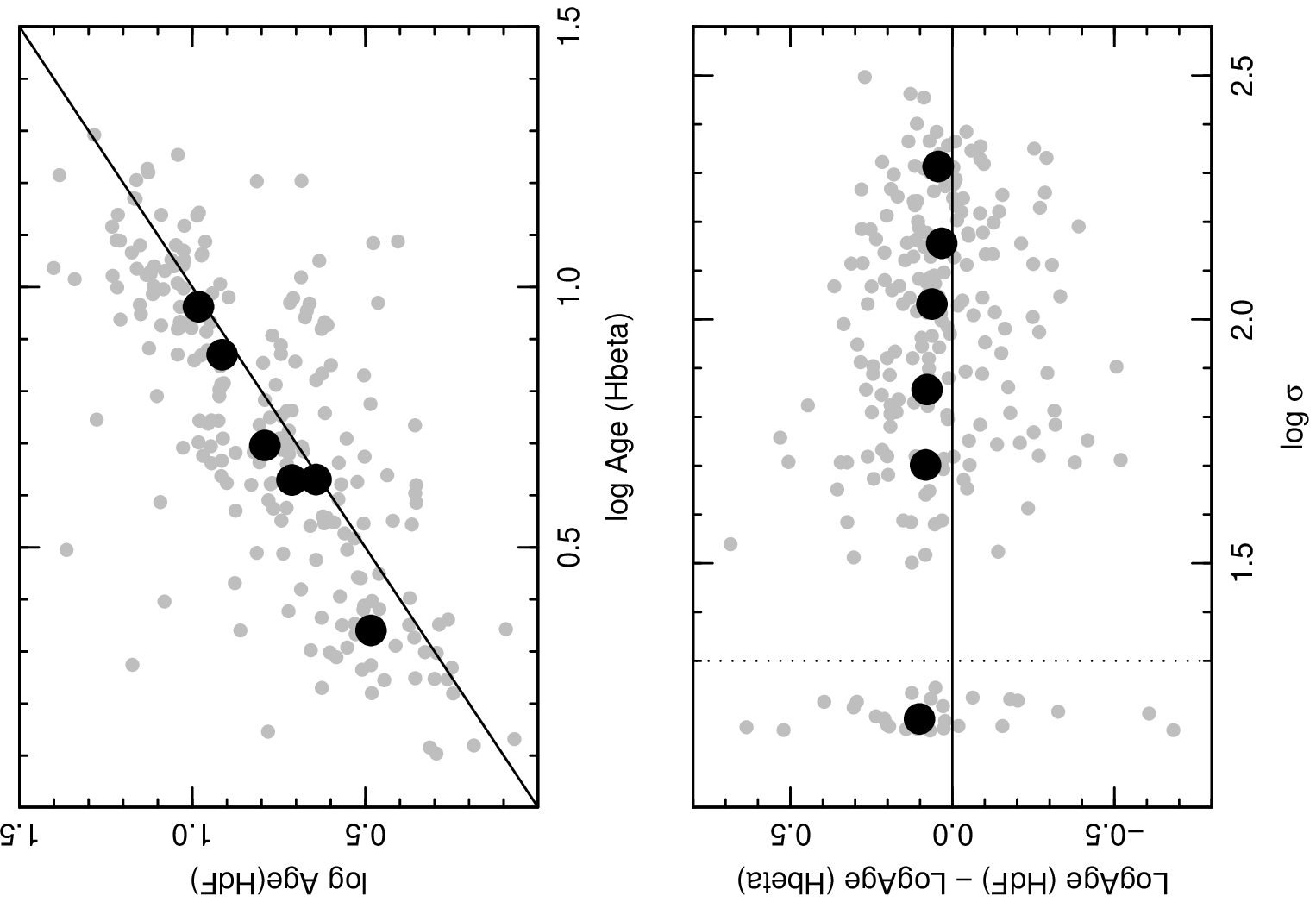}
\caption{
Comparison of age estimates using different Balmer indices. The upper panel shows the estimate based on HdF versus that using Hbeta, while the lower
left plots age difference against $\log\sigma$. 
Small grey points are the individual data; large black points represent medians in $\sigma$ over the same bins as in Figure~\ref{fig:afegrid}.
In the lower panel, galaxies with unresolved velocity dispersion have placed to the left of the dotted line. 
If the younger ages at low mass were due to ``frosting'' by a low-mass-fraction young sub-population, the data would drift to 
lower HdF ages at low $\sigma$. In fact a slight shift in the opposite direction is seen. 
}\label{fig:agecomps}
\end{figure}

\begin{table*}
\caption{Model inversion results. The first three columns give the galaxy identifier, 
log R-band luminosity, log velocity dispersion ($\sigma$ in \kms). 
The following three columns are the log age (in Gyr), 
metallicity (\zh) and $\alpha$-element enhancement (\afe), obtained by inversion of Thomas et al. (2003, 2004) models
using the Mgb5177, Fe4383, Fe5015, HdF, HgF and Hbeta indices. The final three columns give the error correlation
coefficients $C$, from which the covariance matrix on the stellar population parameters can be reconstructed.}
\label{tab:agefits}
\begin{tabular}{lrccccccc}
\hline
Galaxy ID & \multicolumn{1}{c}{$\log\frac{L_R}{L_\odot}$} & $\log\sigma$ & log(Age) & \zh & \afe\ & $C$(age,Z) & $C$(age,$\alpha$) &  $C$(Z,$\alpha$) \\
\hline
NFPJ132328.9-314242 &  9.98 & $2.067\pm0.019$ & $0.753\pm0.048$ & $+0.234\pm0.049$ & $+0.229\pm0.028$ & $-0.85$ & $+0.32$ & $-0.18$ \\ 
NFPJ132330.8-314935 & 10.24 & $2.014\pm0.012$ & $0.557\pm0.030$ & $+0.187\pm0.030$ & $+0.144\pm0.017$ & $-0.85$ & $+0.29$ & $-0.16$ \\ 
NFPJ132335.5-315201 &  9.73 & $2.005\pm0.029$ & $0.646\pm0.067$ & $+0.261\pm0.068$ & $-0.016\pm0.039$ & $-0.86$ & $+0.16$ & $+0.01$ \\ 
NFPJ132337.1-315047 &  9.96 & $2.129\pm0.017$ & $1.041\pm0.035$ & $-0.080\pm0.036$ & $+0.258\pm0.041$ & $-0.76$ & $+0.42$ & $-0.31$ \\ 
NFPJ132345.0-314230 &  9.91 & $1.843\pm0.039$ & $0.469\pm0.060$ & $+0.009\pm0.061$ & $+0.061\pm0.037$ & $-0.93$ & $+0.42$ & $-0.33$ \\ 
NFPJ132348.3-314953 &  9.53 & $1.810\pm0.038$ & $0.702\pm0.062$ & $+0.156\pm0.074$ & $+0.100\pm0.043$ & $-0.85$ & $+0.26$ & $-0.17$ \\ 
NFPJ132355.5-313847 &  9.29 & $1.861\pm0.046$ & $0.867\pm0.102$ & $+0.115\pm0.063$ & $+0.150\pm0.048$ & $-0.74$ & $-0.03$ & $+0.04$ \\ 
NFPJ132406.9-314449 &  9.87 & $2.041\pm0.019$ & $0.985\pm0.039$ & $-0.072\pm0.033$ & $+0.231\pm0.036$ & $-0.76$ & $+0.26$ & $-0.16$ \\ 
NFPJ132412.7-314658 &  9.85 & $1.718\pm0.074$ & $0.454\pm0.072$ & $+0.335\pm0.060$ & $+0.111\pm0.034$ & $-0.87$ & $+0.19$ & $-0.06$ \\ 
NFPJ132418.2-314229 & 10.31 & $2.297\pm0.003$ & $0.901\pm0.028$ & $+0.283\pm0.021$ & $+0.239\pm0.011$ & $-0.86$ & $+0.22$ & $-0.11$ \\ 
NFPJ132423.0-313631 & 10.05 & $2.128\pm0.011$ & $0.779\pm0.040$ & $+0.168\pm0.035$ & $+0.167\pm0.019$ & $-0.87$ & $+0.31$ & $-0.21$ \\ 
NFPJ132425.9-314117 & 10.36 & $2.156\pm0.007$ & $0.825\pm0.048$ & $+0.179\pm0.031$ & $+0.161\pm0.017$ & $-0.83$ & $+0.10$ & $+0.00$ \\ 
NFPJ132426.5-315153 & 10.16 & $2.302\pm0.006$ & $0.874\pm0.037$ & $+0.265\pm0.029$ & $+0.254\pm0.014$ & $-0.87$ & $+0.14$ & $-0.07$ \\ 
NFPJ132431.0-315625 &  9.56 & ---             & $0.438\pm0.082$ & $-0.003\pm0.097$ & $+0.022\pm0.059$ & $-0.94$ & $+0.36$ & $-0.23$ \\ 
NFPJ132440.7-314444 &  9.35 & ---             & $0.595\pm0.085$ & $-0.054\pm0.065$ & $+0.101\pm0.057$ & $-0.88$ & $+0.29$ & $-0.19$ \\ 
NFPJ132507.5-314625 &  9.34 & ---             & $0.729\pm0.086$ & $-0.270\pm0.055$ & $+0.178\pm0.078$ & $-0.87$ & $+0.43$ & $-0.27$ \\ 
NFPJ132514.9-314154 &  9.85 & $2.319\pm0.006$ & $1.192\pm0.032$ & $-0.146\pm0.022$ & $+0.295\pm0.024$ & $-0.88$ & $+0.56$ & $-0.55$ \\ 
NFPJ132517.1-314312 &  9.51 & $1.880\pm0.030$ & $0.602\pm0.050$ & $+0.189\pm0.050$ & $+0.148\pm0.030$ & $-0.86$ & $+0.26$ & $-0.18$ \\ 
NFPJ132520.0-313604 &  9.32 & ---             & $0.307\pm0.045$ & $+0.064\pm0.065$ & $+0.044\pm0.054$ & $-0.92$ & $+0.41$ & $-0.30$ \\ 
NFPJ132526.4-314625 &  9.89 & $2.315\pm0.007$ & $1.076\pm0.016$ & $+0.047\pm0.021$ & $+0.335\pm0.017$ & $-0.72$ & $+0.20$ & $-0.13$ \\ 

\hline
\end{tabular}
\end{table*}

\subsection{Predicted colours: a consistency test}\label{sec:coltest}

In this section, we consider whether the stellar populations inferred from spectroscopy are consistent with the 
broad-band colours observed for the same galaxies. We use $B-R$ colours from the Shapley Optical Survey (SOS)
of Mercurio et al. (2006), measured within a fixed aperture of 4.4\,arcsec diameter. To transform the 
spectroscopic age and metallicity to a predicted colour, we use the linear relation 
\begin{equation} 
(B-R)_{\rm pred} = 1.150 + 0.329 \log({\rm Age}) + 0.297 [{\rm Z/H}]\, , \label{eqn:predbr}
\end{equation}
derived from a fit to Maraston (2005) models with age greater than 1\,Gyr and \zh$>-2$. 
Figure~\ref{fig:modelsos} presents comparisons between observed and predicted colours, demonstrating
the comparable roles played by age and metallicity in reproducing the colours (see also Gallazzi et al. 2006). 

The measured colours exhibit a 1$\sigma$ scatter of 0.09\,mag (estimated directly from the 68\,per cent range of the data, 
to reduce the impact of outliers). 
If only the measured age is used to predict the colour (adopting a common value for the metallicity of all galaxies), 
the residual scatter is reduced to 0.07\,mag. A similar scatter is obtained predicting colour from metallicity and using
a common age for all galaxies. When the measured ages and metallicities are used together to predict the colours, a much 
tighter correlation is recovered, with the residual scatter reduced to 0.04\,mag\footnote{
The most extreme outlier, with $B-R$ some 0.4\,mag bluer than expected for its age and metallicity is NFPJ132802.6--314521. This galaxy was
identified by Miller (2005) as a powerful radio source with probable head--tail morphology, and detected
as an X-ray source by Tajer et al. (2005). Examination of the SOS data confirms that its central colour is indeed unusually blue, 
even compared to its large-aperture colours.}. The (anti-correlated) errors in the age and metallicity estimates account for much of
this scatter, leaving only $\sim$0.02\,mag unexplained.

Our predicted colours do not account for \nafe\ variations, since Maraston's models are for solar abundance ratios. 
Figure~\ref{fig:afemodelsos} shows that the residuals from the predicted vs observed colour relation are mildly anti-correlated 
with measured \nafe. The slope, after rejecting a few outliers, is  $\delta(B-R)/\delta[\alpha/{\rm Fe}]=-0.14\pm0.02$, indicating that 
$\alpha$-enhanced populations are slightly bluer than solar-mixture populations. This result is qualitatively consistent with 
theoretical modelling by Coelho et al. (2007), who predict a slightly larger colour change 
$\delta(B-R)/\delta[\alpha/{\rm Fe}] \approx -0.2$\,mag, when Z/H is held 
constant\footnote{Coelho et al. tabulated colour responses to \nafe\
variations at fixed Fe/H. When these are transformed to responses at fixed Z/H, the increase in \nafe\ is accompanied by a reduction
in Fe/H; both effects generate bluer colours.}. Allowing for the observed dependence on \nafe, and for measurement errors, the colours show 
a residual intrinsic scatter of only 0.01\,mag, compared to predictions from the spectroscopic parameters.

\subsection{Constraints on composite stellar populations}\label{sec:compo}

We have so far been comparing observed galaxy spectra to idealised simple, i.e. single burst, stellar population (SSP) models. 
In a hierarchical galaxy formation model, real star-formation histories (SFHs) are expected to be extended, especially
before incorporation into a cluster, and perhaps to have multiple bursts as new gas becomes available from mergers. 
In this case, we can still fit for an SSP-equivalent population, which best matches the observed parameters, 
but we need to be very careful regarding its interpretation. In general, the SSP-equivalent age is not a good approximation
to the luminosity-weighted age. Rather, because the Balmer lines respond non-linearly with age, the 
SSP-equivalent age is even more heavily weighted to the youngest stars present. An excellent discussion of these
issues, with extensive simulations of two-burst  SFHs, is given by Serra \& Trager (2006); the problem was
discussed previously by Leonardi \& Rose (1996) and Trager et al. (2000), among others. In this section, we briefly assess the
evidence for composite stellar populations in the Shapley supercluster sample. 
Allanson et al. (2009) provide a more general analysis of our data using complex SFH models. 

Using line indices alone, it is very difficult to distinguish between complex histories which have no star-formation in 
the past $\sim$1\,Gyr, because changes in the spectrum are fairly linear in this regime. As a result, two widely separated
star-formation events can produce spectra that are indistinguishable from a single burst with intermediate age. 
For such cases, a single SSP-equivalent age still provides a convenient moment of the age distribution, that
can be readily compared to galaxy formation models (e.g. Smith et al. 2009a; Trager \& Somerville 2009). 
By contrast, sub-populations younger than 1\,Gyr show spectral signatures of A-stars, 
which cannot be mimicked by intermediate-age SSPs. 

As a test for ``frosting'', i.e. young sub-populations representing low fractions of the total mass, 
we can compare the SSP-equivalent ages determined using different age indicators. 
The luminosity contribution from any younger sub-population would be stronger at \hd\ than at \hb, for example, 
due to the bluer colour of their main-sequence turn-off stars. 
As a result the \hd-derived ages would be younger than the \hb-derived ages. In Figure~\ref{fig:agecomps}, we compare
the ages derived from an inversion using Hbeta--Fe5015--Mgb5177 to those from  HdF--Fe5015--Mgb5177. The figure shows that 
the HdF ages are in fact {\it older} than the ages based on Hbeta, by $\sim$0.07\,dex, the opposite of the offset expected from
frosting. Moreover the offset is larger for low$-\sigma$ galaxies than for the most massive objects, so an increased incidence of frosting
is apparently not the cause of the overall trend towards younger age at low mass. (This is also seen in the fits in Paper I, where the 
slope of the age--$\sigma$ relation is steeper for the fits based on Hbeta.)

An alternative indicator for young sub-populations is the Rose (1984) \caii\ index, based on the residual flux 
at the core of the Ca K and H lines, the latter being coincident with the higher-order Balmer line H$\epsilon$. The index has a very non-linear behaviour, being
approximately constant for all stars cooler than spectral type F2, but changing steeply for hotter stars.
As a result, in composite populations, \caii\ is very sensitive to the presence stars of $\la$1\,Gyr age
 even if they contribute only 
a few percent of the mass, but little affected by variations in age beyond this threshold. 
In combination with other Balmer indices,
the \caii\ index can distinguish intermediate-age approximate SSPs from old-plus-young composite populations (Leonardi \& Rose 1996).

\begin{figure}
\includegraphics[angle=0,width=85mm]{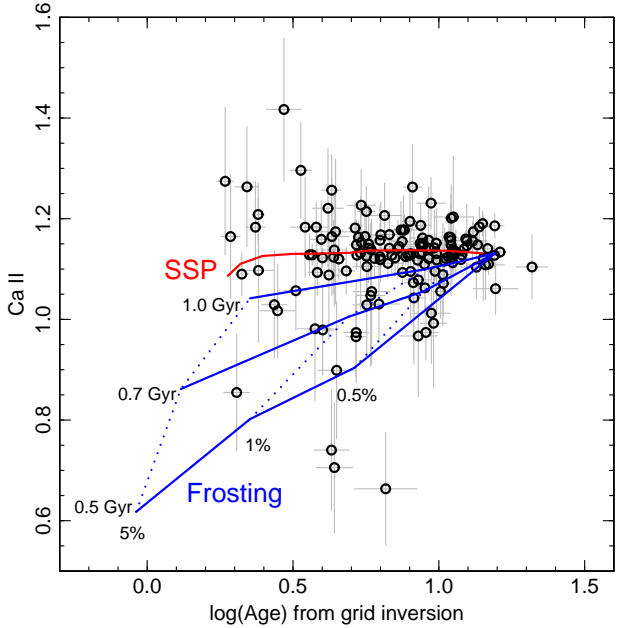}
\caption{
The Rose \caii\ index for galaxies with errors smaller than 0.15, compared to the SSP-equivalent age. The blue grid indicates
the expected behaviour for frosting by secondary bursts of age 1.0, 0.7 and 0.5\,Gyr (solid lines), and mass-fractions
5, 1 and 0.5\,per cent (dotted lines), with the remaining mass in a 13\,Gyr base population. The red track shows predictions for 2--12\,Gyr SSPs, 
demonstrating the stability of \caii\ for intermediate ages. Although a few galaxies fall in the region of the frosted models, {\it on average} the SSP-equivalent ages of
``young'' galaxies are not driven by secondary bursts in the past Gyr. 
}\label{fig:rosecaii}
\end{figure}

In Figure~\ref{fig:rosecaii}, we show \caii\ versus the age derived from our default fits.
For clarity we show only the 149 galaxies (64 per cent of the sample) with errors less than $0.15$ in \caii. 
For comparison we show predictions for \caii\ for SSPs and for populations with a frosting of
young stars, derived by combining SSP models from Vazdekis et al. (2009). The cases shown are for a $\sim$13\,Gyr base population, 
with a secondary burst ages of 0.5, 0.7 and 1.0\,Gyr. The mass fraction in the burst is 0.5, 1.0 and 5.0\,per cent. Both populations
are assigned solar Fe/H (metallicity effects on \caii\ are small enough to neglect for this test). For the 
horizontal axis, we compute the SSP-equivalent ages for the composite two-burst spectrum, by inverting the TMBK
models using the same six indices that were used for the real data. (The mixing of different model sets is clearly undesirable, 
but cannot be avoided for this test; the results are probably fairly generic.) These simulations predict
easily-observable \caii\ offsets for galaxies with SSP-equivalent ages $\la5$\,Gyr, if their ages are driven by minor 
bursts within the last $\la$1\,Gyr. It is seen that a few galaxies indeed lie in the region occupied by the frosting models, and
inconsistent with the cool-star \caii\ value of $\approx$1.1. However, the overall distribution of points indicates that frosting is not
responsible for the 
young derived ages on average\footnote{
If the galaxies with larger \caii\ errors are included, the fraction of frosting candidates remains small. This is expected, since
frosted populations have more flux in the blue, and are thus over-represented in the low-error subset.}.
In particular, for 22 galaxies with inversion ages less than 4\,Gyr, the mean
\caii\ is 1.132$\pm$0.026, indistinguishable from the value 1.135$\pm$0.005 for 40 galaxies with inversion ages above 10\,Gyr 
(the median values are 1.128 and 1.134 respectively). 

In conclusion, we find no evidence that SSP ages are {\it on the whole} generated by frosting by small mass-fractions 
of very recent star-formation, although we cannot exclude this possibility for a small
number of individual sample galaxies. The same conclusion holds for any other contribution from A-type stars,
such as blue Horizontal Branch stars from an old, metal-poor sub-population (e.g. Maraston \& Thomas 2000). 
Allanson et al. (2009) show that frosting models (or exponential decay models, which similarly allow star-formation within the 
past Gyr) are inconsistent with the dynamical mass-to-light ratios implied by the Fundamental Plane.

\subsection{Corrections for stellar population gradients}\label{sec:apcorrs}

The 2\,arcsec diameter fibres enclose a larger fraction of the total light for small galaxies than for large ones. 
For galaxies in our sample, the effective (i.e. half-light) radii ($R_{\rm e}$) range from 1\,arcsec to 10\,arcsec
(Gargiulo et al. 2009).  Since red galaxies in general exhibit internal metallicity gradients (e.g. Davies, Sadler \& Peletier 1993), 
with higher metallicity in the central regions, a correction is needed to translate the central measurements to estimates of the 
metallicity at a common radius in terms of $R_{\rm e}$. Neglecting this correction could lead to systematic trends that can
mimic correlations with galaxy mass. 

\begin{figure}
\includegraphics[angle=270,width=85mm]{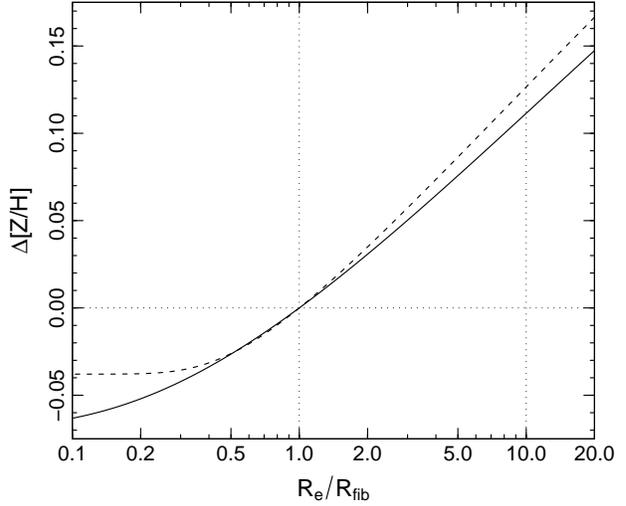} 
\caption{Aperture effects on fibre measurements of \zh. The curves shows the difference between \zh\ measured within a
circular fibre of radius $R_{\rm fib}$ and that measured at the effective radius, $R_{\rm e}$, as a function of the ratio between these radii.
The solid curve is for an $R^{1/4}$ luminosity profile, and the dashed curve is for an exponential profile. We assume a differential 
metallicity gradient of 0.14 dex per decade in radius. For our Shapley sample, the effective radii are in the range 1.0--10.0\,arcsec,and
$R_{\rm fib}$=1\,arcsec. Over this range, indicated by the vertical dotted lines, a linear approximation to the aperture correction is justified. }
\label{fig:apeffect}
\end{figure}

To estimate the required corrections, we will assume metallicity gradients of order $\Delta[{\rm Z/H}]/\Delta\log R=-0.14$ (Rawle et al. 2008b), 
for all galaxies. Weighted by an $R^{1/4}$ luminosity profile, and integrated within circular apertures, this yields a 
correction of $\sim$0.1\,dex in Z/H between a measurement within  $0.1\,R_{\rm e}$ and what would be observed within 
1\,$R_{\rm e}$ (Figure~\ref{fig:apeffect}). 

Accounting for metallicity gradients is especially important when we attempt to disentangle the effects
of luminosity and velocity dispersion on the stellar populations.  The Fundamental Plane implies that at fixed
velocity dispersion, more luminous galaxies are larger. In turn this means the offset between metallicity at $R_{\rm e}$
and metallicity within the fibre is larger for more luminous galaxies. Conversely, at fixed luminosity, low-$\sigma$ galaxies
are larger than high-$\sigma$ galaxies. 
If uncorrected, metallicity gradients would thus artificially boost the correlation of Z/H with luminosity at fixed 
$\sigma$ and suppress its correlation with $\sigma$ at fixed luminosity. 

Effective radii have not been measured for all of our sample galaxies, but for the subset of 141 objects studied by Gargiulo et al.
we find 
\[
R_{\rm e} = 2\,\farcs15 \left(\frac{L_R}{10^{10}L_\odot}\right)^{+1.03}\left(\frac{\sigma}{100\,{\rm km\,s}^{-1}}\right)^{-1.09}\,,
\]
with a 1$\sigma$ scatter of 0.13\,dex.
Applying this relation (simply a statement of the Fundamental Plane) to calculate a predicted effective radius $R_{\rm e}^{\rm pred}$ for all galaxies 
in our sample, we can correct the metallicity to $R_{\rm e}^{\rm pred}$ via
\[
\Delta{\rm [Z/H]} = -0.1 \,\log(R_{\rm e}^{\rm pred}/R_{\rm fib})\,.
\]
The corrections to Z/H range from $-0.09$\,dex to $+0.06$\,dex, with median $-0.02$\,dex\footnote{
Note that within the linear part of the correction curve (Figure~\ref{fig:apeffect}), 
the {\it range} of aperture corrections depends only on the range of effective radius or equivalently
the range of luminosity, if considering fixed $\sigma$. It is independent of the actual 
fibre size, which affects only the {\it average} correction. 
}.
The error in the 
the correction, associated with scatter around the Fundamental Plane, is 0.1$\times$0.13\,$\approx$\,0.01\,dex, which is much smaller than the formal 
error on Z/H (typically $\sim0.05$\,dex), and is neglected in what follows. 
The strength of the internal gradients, i.e. the factor $-0.1$ above, is uncertain to perhaps 50 per cent. 
In Section~\ref{sec:agemass}, we consider explicitly how our results for Z/H would change if the adopted aperture corrections
are altered within this range. 

The evidence for internal age gradients in red galaxies is inconclusive, with small positive (Rawle et al. 2008b) and negative
(\sanch\ et al. 2006b) gradients having been reported. Similarly, gradients in \nafe\ appear to be small. No corrections are
imposed here for either parameter. For the velocity dispersions, the usual correction is adopted, with
\[
\log\frac{\sigma(R_{\rm e}^{\rm pred})}{\sigma(R_{\rm fib})} = -0.04 \log\frac{R_{\rm e}^{\rm pred}}{R_{\rm fib}} 
\]
(e.g. J\o rgensen, Franx \& Kj\ae rgaard 1995). The corrections range from $-0.04$\,dex to $+0.02$\,dex in $\sigma$.

\section{Observed scaling relations of the stellar populations}\label{sec:agemass}

\begin{table}
\caption{
Observed scaling relation fits, adopting the default aperture correction model. For each parameter Z/H, age, and \nafe, we provide fits with
velocity dispersion $\sigma$, R-band luminosity $L_R$, and to both parameters simultaneously. In fits to the residuals
from one of the relations (indicated by $\Delta$), the fixed coefficient is given in parentheses. 
The tabulated scatter is the estimated {\it intrinsic} dispersion around the fit. 
The first line in each section gives the scatter around the null model, e.g. the total
metallicity scatter of the sample as a whole. 
}
\label{tab:fitcoeffswkap}
\begin{tabular}{lccc}
\hline
Relation & Coeff of $\log\sigma$ & Coeff of $\log{}L_R$ & scatter \\
\hline
\zh\       &  ---              &  ---              &  0.151 \\
\zh\ -- $\sigma$       &  +0.307$\pm$0.043 &  ---              &  0.133 \\
\zh\ -- $L_R$            &  ---              &  +0.216$\pm$0.028 &  0.131 \\
$\Delta$\zh\ -- $\sigma$ &  +0.067$\pm$0.042 &  (+0.216)         &  0.130 \\
$\Delta$\zh\ -- $L_R$     &  (+0.307)         &  +0.064$\pm$0.028 &  0.131 \\
\bf{\zh\ -- $\sigma,L_R$}     &  \bf{+0.149$\pm$0.062} &  {\bf +0.143$\pm$0.042} &  {\bf 0.129} \\
\hline
log(Age)       &  ---              &  ---              &  0.194 \\
log(Age) -- $\sigma$       &  +0.437$\pm$0.053 &  ---              &  0.165 \\
log(Age) -- $L_R$            &  ---              &  +0.112$\pm$0.041 &  0.191 \\
$\Delta$log(Age) -- $\sigma$ &  +0.312$\pm$0.056 &  (+0.112)         &  0.176 \\
$\Delta$log(Age) -- $L_R$     &  (+0.437)         &  --0.106$\pm$0.035 &  0.161 \\
{\bf log(Age) -- $\sigma,L_R$}     &  \bf +0.698$\pm$0.075 & \bf --0.235$\pm$0.050 & \bf  0.156 \\
\hline
\afe\       &  ---              &  ---              &  0.091 \\
\afe\ -- $\sigma$       &  +0.210$\pm$0.026 &  ---              &  0.076 \\
\afe\ -- $L_R$            &  ---              &  +0.047$\pm$0.020 &  0.089 \\
$\Delta$\afe\ -- $\sigma$ &  +0.156$\pm$0.027 &  (+0.047)         &  0.081 \\
$\Delta$\afe\ -- $L_R$     &  (+0.210)         &  --0.053$\pm$0.017 &  0.074 \\
\bf \afe\ -- $\sigma,L_R$     &  \bf +0.341$\pm$0.035 &  \bf --0.117$\pm$0.024 & \bf 0.070 \\
\hline
\end{tabular}
\end{table}

\begin{table}
\caption{Effect of varying the aperture correction scheme on the coefficients obtained for \zh\ as a function of $\sigma$ and $L_R$.}
\label{tab:apcorrtests}
\begin{tabular}{lccc}
\hline
Scheme & $\frac{\Delta {\rm Z/H}}{\Delta\log R}$ & Coeff of $\log\sigma$ & Coeff of $\log{}L_R$ \\
\hline
None     & \phantom{--}0.00  &  +0.044 &  +0.241  \\
Weaker   &           --0.07  &  +0.097 &  +0.192  \\
{\bf Default} & {\bf --0.15} &  \bf{+0.149} &   {\bf+0.143} \\
Stronger &           --0.23  &  +0.201 &  +0.093 \\
\hline
\end{tabular}
\end{table}

In this section we examine the dependence of age, Z/H and \nafe\ as functions of velocity dispersion, $\sigma$,
and luminosity, $L_R$.  We start with one-parameter descriptions, e.g. the age$-\sigma$ relation, 
then test for residual correlations, and finally consider bivariate fits in which $\sigma$ and $L_R$ are used 
simultaneously to predict the stellar populations. 

We defer a discussion of the correlation {\it among} stellar population parameters, e.g. the age--metallicity 
and  age--\nafe\ relations, to a subsequent paper in this series. 

\subsection{Biases and choice of the mass proxy}\label{sec:proxychoice}

Velocity dispersion has traditionally been the mass proxy used most widely for stellar population studies, 
because it is independent of the stellar mass-to-light ratio and reflects the depth of the gravitational
potential well. However, when galaxy samples are {\it selected} by luminosity, the use of $\sigma$
as a mass proxy is biased: with decreasing $\sigma$, such samples become increasingly dominated by galaxies with 
unusually high luminosity given their $\sigma$.

The most obvious mechanism for such a bias is that old galaxies are fainter than otherwise identical
galaxies with younger populations (because stellar mass-to-light ratio increases with age)
 and are hence lost at low velocity dispersion in a luminosity-selected sample
(as discussed by Kelson et al. 2006, for example). If not corrected for, this generates an artificial steepening
of the age$-\sigma$ relation. To a lesser extent, the same effect steepens the Z/H$-\sigma$ relation, since the
stellar mass-to-light ratio also increases slightly with metallicity. 

A second cause of bias can be identified, which is distinct from the above. Consider a parameter $P$, which is
positively  correlated both with stellar mass (and hence luminosity), and with $\sigma$. If we construct the $P-\sigma$ relation
from a luminosity-selected sample, the observed low-$\sigma$ galaxies are unrepresentative of the population
at large, being unusually bright for their $\sigma$. In turn, they have unusually high values of $P$, 
and the slope of the $P-\sigma$ relation will be artificially flattened. 
This is irrespective of any direct causality between $P$ and $L_R$. 
A bias of this type was considered by Graves et al. (2007) when deriving composite spectra for SDSS galaxies
binned by velocity dispersion. They found that within a given $\sigma$ interval, certain metal-line indices 
showed a strong correlation with luminosity. As a result, at low $\sigma$ their sample was biased to objects
with unrepresentatively high metallicity, and they introduced a correction method to account for the bias. 
In the following section, we will instead include the apparent luminosity dependence explicitly in our
analysis of the scaling relations.

\subsection{Correlations with luminosity and velocity dispersion}\label{sec:lumsigscals}

We begin by considering qualitatively the distribution of measured stellar population parameters, and their correlation with either
the R-band luminosity $L_R$ or with velocity dispersion $\sigma$. We use only the set of 193 galaxies which have 
reliable stellar population fits and resolved velocity dispersions. When fitting the correlations, we weight galaxies
according to the quadrature sum of their measurement errors and the intrinsic scatter in the fit.  In practice, this
is done by iteratively re-weighting the points, updating the estimate of intrinsic scatter in each step. 
The errors in predictors ($\sigma$, $L_R$) are ignored in the fits. Table~\ref{tab:fitcoeffswkap} presents the coefficients
of the fits described in this section. 

\begin{figure*}
\includegraphics[angle=270,width=180mm]{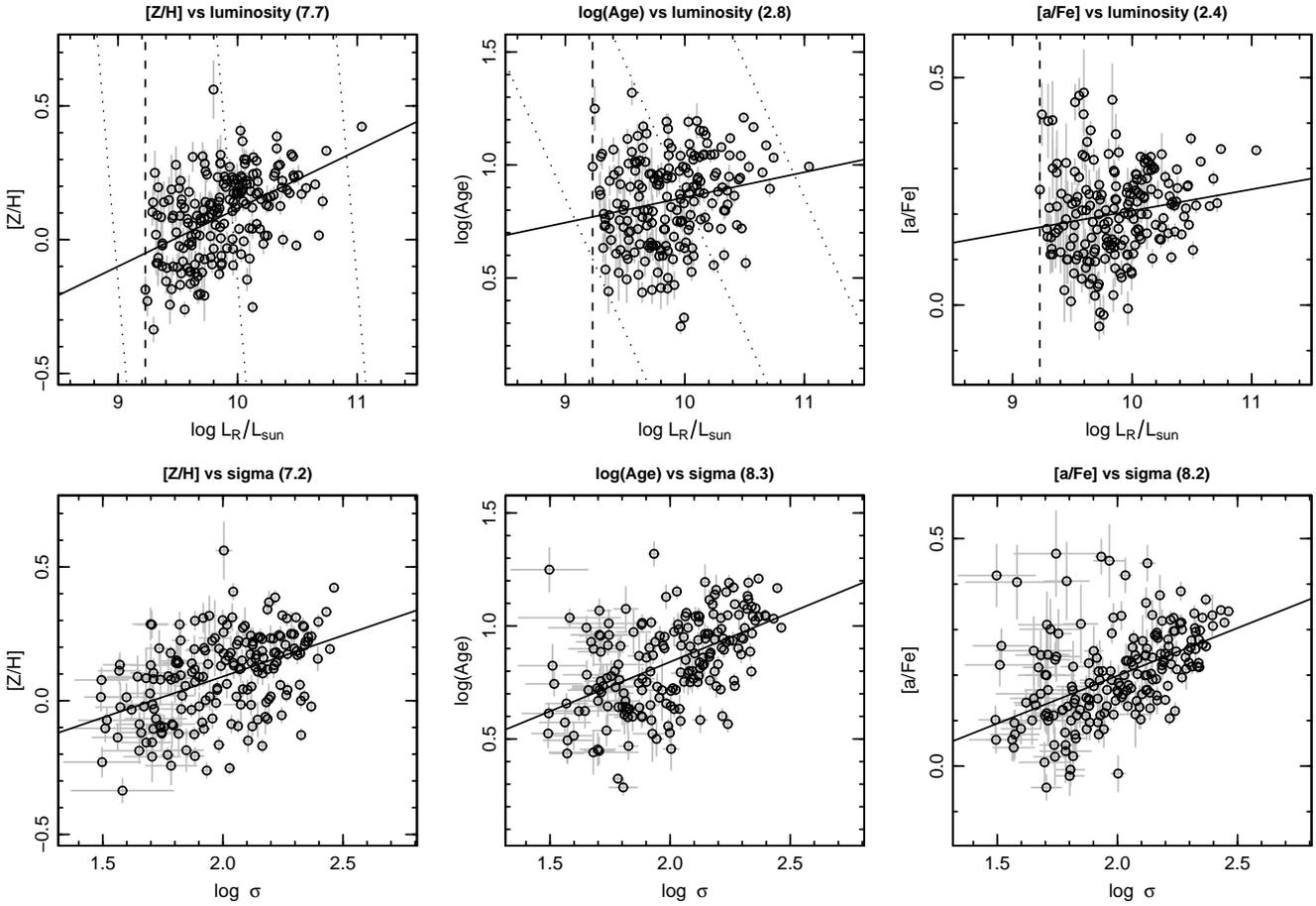} 
\caption{Stellar parameters versus two alternative mass proxies.
The upper panels show the trends with luminosity, i.e. the  Z/H$-L_R$, age$-L_R$ and \nafe$-L_R$ relations. 
The sample selection limit is indicated by the vertical dashed line. 
For the Z/H panel, lines of constant stellar mass 
($M_* = 10^{9.5}, 10^{10.5}, 10^{11.5} M_\odot$)
are shown by dotted lines,
assuming an age of 10\,Gyr. In the age panel, solar metallicity is assumed for the stellar mass lines. 
In the lower panels, we show the equivalent trends with velocity dispersion $\sigma$. 
The horizontal axis scales in the lower panels have been 
matched to the upper panels using the relationship between $L_R$ and $\log\sigma$ for this sample. 
The significance of the fitted slope (in units of the standard error) is indicated in parentheses in the title-bar. 
For the coefficients of the fits, see Table~\ref{tab:fitcoeffswkap}.
}\label{fig:fullcorplots}
\end{figure*}

The correlations with luminosity are shown in the upper panels of Figure~\ref{fig:fullcorplots}. 
The metallicity Z/H exhibits a very strong ($\sim$8$\sigma$) correlation with $L_R$, while age and
\nafe\ are more weakly correlated ($<3\sigma$).
Because the sample is luminosity-limited, the selection cut is vertical in these diagrams, and 
does not explicitly bias the slopes of the relationships. However, 
because stellar mass-to-light ratios depend on age, and to a lesser extent
on metallicity, lines of constant stellar mass run diagonally across the panels. 
In the lower panels of Figure~\ref{fig:fullcorplots}, we show the correlations with central velocity 
dispersion. The Z/H$-\sigma$ relation is slightly weaker than the Z/H$-L_R$ relation ($\sim$7$\sigma$), 
while the relations for age and \nafe\ are much steeper ($\sim$8$\sigma$) than the equivalent 
$L_R$ correlations. 
The trends with $\sigma$ are in good agreement with the estimates made in Paper I based on the slopes of the index--$\sigma$
relations (i.e. Z/H\,$\propto$\,$\sigma^{0.34}$, age\,$\propto$\,$\sigma^{0.52}$, $\alpha$/Fe\,$\propto$\,$\sigma^{0.23}$ from table~7 of Paper I), 
although the new age slope is slightly shallower. 
We note that the fit to the \nafe$-\sigma$ relation seems to be flattened by the 
presence of high-\nafe\ outliers at relatively low $\sigma$. The ``core'' of the relation would favour a slope 
$\sim$50\,per cent steeper, which would be in closer agreement with previous work (slopes of 0.3--0.4 were derived by 
Nelan et al. 2005, Thomas et al. 2005,  Bernardi et al. 2006, Graves et al. 2007). The fairly shallow \nafe$-\sigma$ relation
derived in Paper I is possibly influenced in the same way. 

\begin{figure*}
\includegraphics[angle=270,width=180mm]{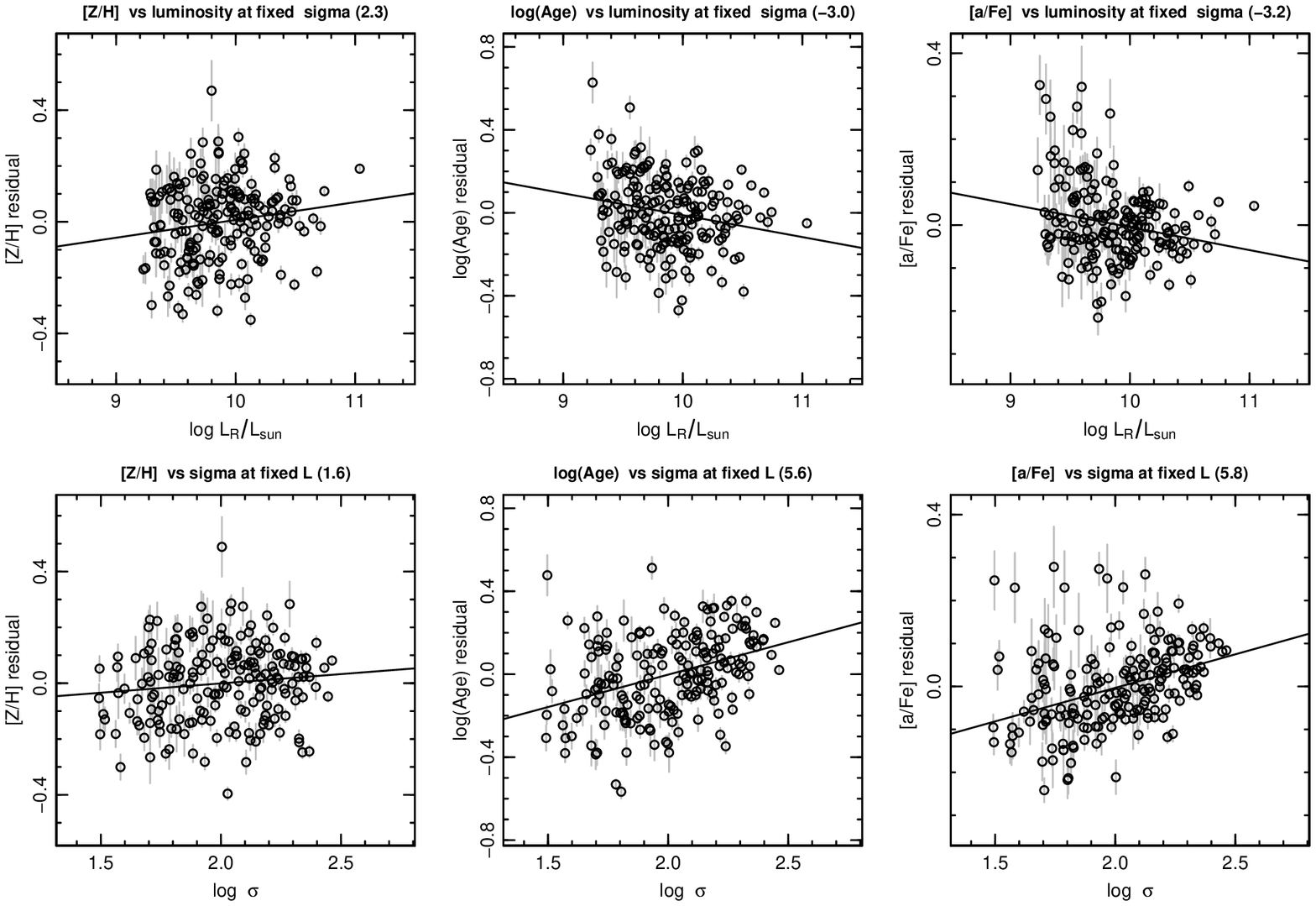} 
\caption{Residual correlations with the mass proxies. In the upper row, we show the correlations
of age, Z/H and \nafe\ with luminosity, after removing the trends with velocity dispersion. In the 
lower panel, the trends with $\sigma$ are shown after removing the luminosity trends. Note the different
behaviour of Z/H compared to the other two parameters. The significance of the fitted slope (in units of the standard error) is indicated in parentheses in the title-bar.
For the coefficients of the fits, see Table~\ref{tab:fitcoeffswkap}.
}\label{fig:partcorplots}
\end{figure*}

Figure~\ref{fig:partcorplots} shows the residual correlations. 
In the upper panels, we plot the residuals from the 
Z/H$-\sigma$, age$-\sigma$, and \nafe$-\sigma$ relations, as a function of luminosity. 
Fairly weak residual luminosity trends ($2-4$$\sigma$) are found for all three parameters: 
Z/H increases with luminosity at fixed $\sigma$, 
while age decreases with luminosity at fixed $\sigma$; \nafe\ also formally decreases with luminosity, although
the ``core'' of the distribution is fairly flat. 
The lower panels of Figure~\ref{fig:partcorplots} show, conversely, the residual dependence on $\sigma$, 
after controlling for luminosity. The correlation of Z/H  with $\sigma$ is reduced to $<$2$\sigma$ significance
after accounting for its variation with $L_R$. On the other hand, the age and \nafe\ correlations with 
$\sigma$ are strengthened when we control for the luminosity trend. These tests confirm the 
indications from Figure~\ref{fig:fullcorplots} that velocity dispersion is dominant in driving 
age and \nafe, while Z/H depends at least partly on luminosity.

To explore these correlations further, we next consider bivariate scaling relations, where
age, Z/H and \nafe\ are expressed explicitly in terms of both  $L_R$ and $\sigma$, 
i.e. $P = a \log\sigma + b \log(L_R) + c$, with $P$ = log(Age), \zh, \afe. This formulation is unbiased by selection
in luminosity, and also by selection in velocity dispersion (recall that some 15\,per cent of the Shapley galaxies are ``unresolved'', indicating
that such a selection does affect our sample). The robustness holds so long the distribution of galaxy properties at given $\sigma$ 
{\it and} given $L_R$ is not affected by additional selection criteria. 

\begin{figure*}
\includegraphics[angle=270,width=180mm]{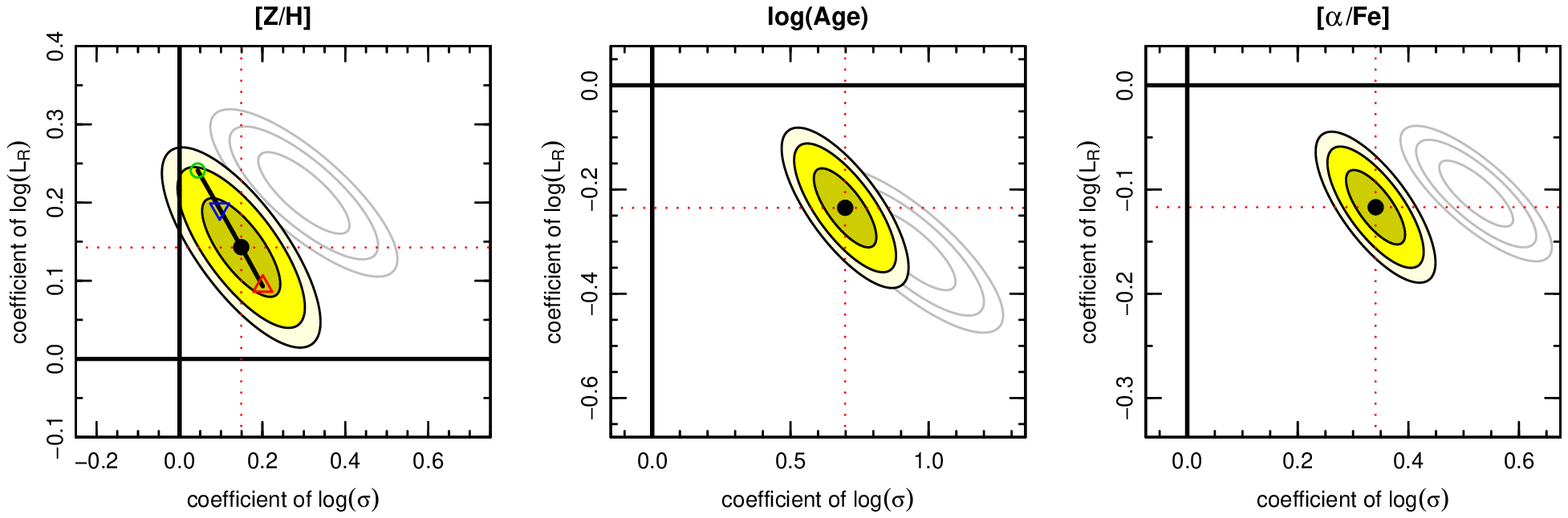} 
\caption{Bivariate model coefficients. The shaded ellipses show the 1,\,2,\,3\,$\sigma$ intervals on the coefficients of fits
log(Age)=$a_0 + a_1 \log\sigma + a_2 \log L_R $, and similarly for
\zh\ and \afe. 
For the metallicity panel, triangles indicate the fits obtained adopting 50\,per cent stronger (red, upward-pointing)
and  50\,per cent weaker (blue, downward-pointing) aperture corrections for metallicity gradients (see Table~\ref{tab:apcorrtests}). The open circle shows the fit
when no aperture correction is applied at all. 
Because the predictors $\sigma$ and $L_R$ are correlated, the errors on their 
coefficients are anti-correlated. 
The unshaded grey ellipses indicate the correlations obtained from fitting the SDSS data from Graves et al. (2009), 
as discussed in Section~\ref{sec:discuss}.
} \label{fig:lumsigfit} 
\end{figure*}

\begin{figure*}
\includegraphics[width=61mm,angle=270]{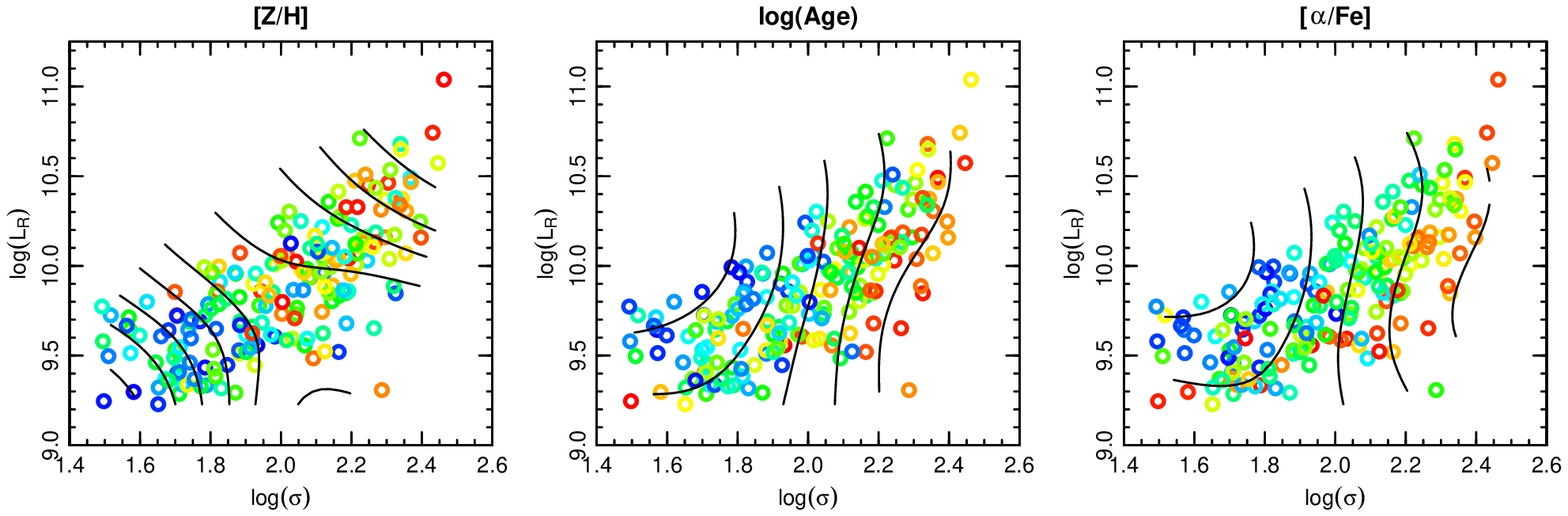}  
\caption{Marked Faber--Jackson relations, showing the distribution of SSP parameters in the space of
luminosity and velocity dispersion. The colours are assigned according to {\it rank} in each parameter shown.
The contours show lines of equal parameter rank, with a heavy smoothing applied.}
\label{fig:markfjpars}
\end{figure*}

We present the coefficients of the bivariate fits in Table~\ref{tab:fitcoeffswkap}, and illustrate the results
graphically in Figure~\ref{fig:lumsigfit}. 
For both age and \nafe, the bivariate fits confirm a positive correlation with $\sigma$ and simultaneously a negative correlation
with luminosity. For Z/H, there are positive dependences on both $\sigma$ and $L$. Table~\ref{tab:apcorrtests} shows that
the luminosity correlations are sensitive to the adopted aperture corrections. If no correction at all is allowed
for the metallicity gradients, the recovered trend of Z/H would be primarily with luminosity. 
An alternative way to present the dependence of parameters on luminosity and velocity dispersion is through ``marked Faber--Jackson'' relations 
(Figure~\ref{fig:markfjpars}) in which the location in the $(L,\sigma)$ plane is colour-coded by age, Z/H, \nafe\ in turn. These figures
provide an immediate visual confirmation of the different behaviour of Z/H (inclined contours) relative to the other parameters (contours aligned
mainly with $\sigma$). A somewhat comparable illustration, for age, is shown by Gallazzi et al. (2006) in their figure 15a. 

\subsection{Bivariate scaling relation for the broadband colour}\label{sec:bivindex}

Using the bivariate scalings of age and metallicity with velocity dispersion and luminosity, it is now possible to predict the 
equivalent correlations for broadband colours, and compare to the observed trends. 

A bivariate fit to the observed $B-R$ colours from SOS, within an aperture of 
4.4\,arcsec diameter, yields a correlation primarily with velocity dispersion:
$\Delta(B-R)_{\rm obs} = (0.19\pm0.03) \Delta\log\sigma - (0.01\pm0.01) \Delta\log{}L$. 
This agrees with the findings of Bernardi et al. (2005) who showed that a strong colour$-\sigma$ 
relation holds after controlling for luminosity, but that
there is no colour--luminosity relationship at fixed $\sigma$ (their Figure~4).

We can readily explain this behaviour using our spectroscopically-determined age and metallicity scaling relations. 
Since the observed colours are within a fixed aperture, we use the scalings derived for the ``no aperture correction''
case. The factor-of-two mismatch between the photometry aperture and spectrograph fibre size should yield only a constant
correction factor, and will not affect the slopes of the correlation. 
Using Equation~\ref{eqn:predbr}, which was derived from the Maraston (2005) models, predicted colour 
differences can be expressed as $\Delta(B-R)_{\rm pred} = 0.33\Delta\log({\rm Age}) + 0.30\Delta[{\rm Z/H}]$. 
The small effects of \nafe\ on broadband colours will be neglected here. 
Inserting the $L,\sigma$ dependences of age and metallicity, this gives
$\Delta(B-R)_{\rm pred} = 0.23 \Delta\log\sigma - 0.08\Delta\log{}L$ from the age change and 
$\Delta(B-R)_{\rm pred} = 0.01 \Delta\log\sigma + 0.07\Delta\log{}L$ from Z/H variations. 
We see that the age and metallicity contributions to the $L$ dependence cancel almost precisely, 
leaving $\Delta(B-R)_{\rm pred} = 0.24 \Delta\log\sigma - 0.01\Delta\log{}L$, dominated by the change in $\sigma$, 
as observed. 

This test confirms the consistency between the derived spectroscopic parameters and the observed broadband colours. 
Where Section~\ref{sec:coltest} demonstrated this on a galaxy-by-galaxy basis, here we have shown that the derived 
systematic behaviour is also reflected in the colours.

\section{No correlations with stellar mass?}\label{sec:stelmass} 

\begin{figure*}
\includegraphics[angle=270,width=180mm]{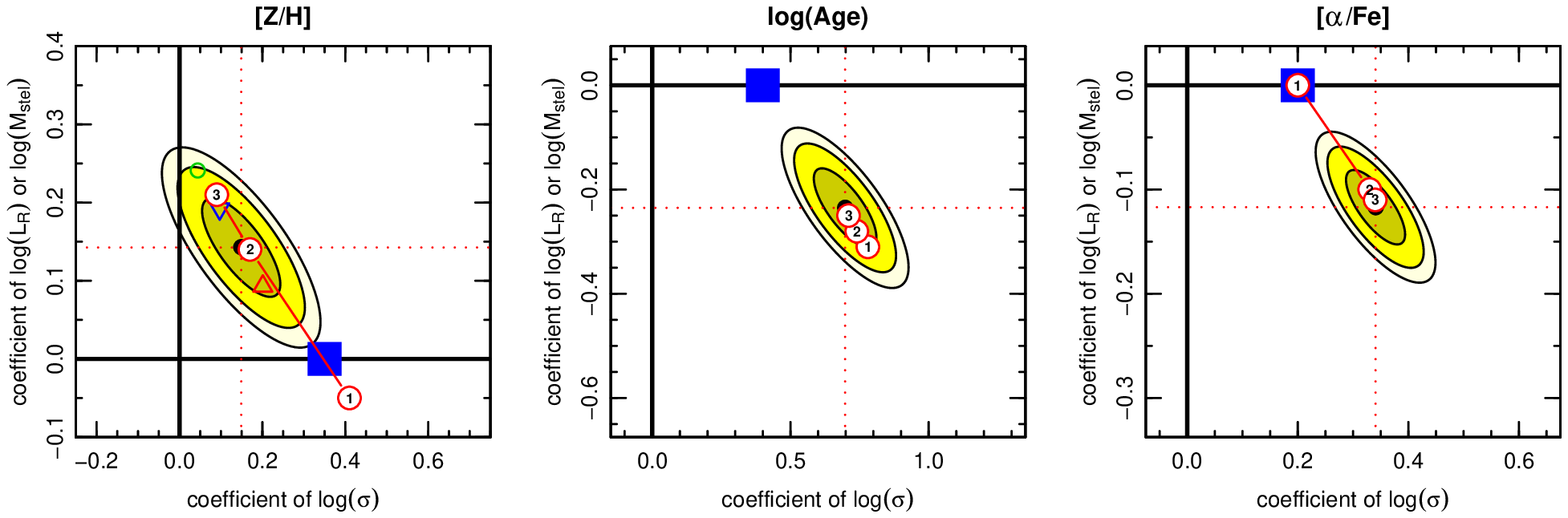} 
\caption{The observed correlations (yellow ellipses, as in Figure~\ref{fig:lumsigfit}) compared to the outputs from simulations that assume
no intrinsic dependence on stellar mass. The velocity dispersion dependence of the input models is indicated by the blue
square on the horizontal axis. The correlations with $\sigma$ and $L$, obtained by fitting the simulated dataset, are shown by the numbered red circles, 
for three cases described in the text: 1 for uncorrelated age/metallicity scatter; 2 for moderately anti-correlated scatter; 3 for perfectly anti-correlated
scatter. The figure demonstrates that the observed luminosity dependences can be recovered without any intrinsic correlation with stellar mass, provided
that the age/metallicity scatter is anti-correlated. 
} \label{fig:lumsigfit2} 
\end{figure*}

\begin{table*} 
\caption{Results for Monte Carlo modelling of the underlying scaling relations. The input model assumes that the stellar populations depend intrinsically only on velocity dispersion ($\sigma$),
and are not systematically related to stellar mass ($M_*$). The recovered correlations with $\sigma$ and luminosity ($L_R$) are comparable to the observed relations, when correlated
residuals (the Z-plane) are allowed for. Solutions (1), (2) and (3) are as labelled in Figure~\ref{fig:lumsigfit2}}\label{tab:massfits}
\begin{tabular}{llllll}
\hline
Input model $\sigma,M_*$ scalings              & ${\rm Z/H}\,\propto\sigma^{+0.35}M_*^{\,0.00}$ &  ${\rm Age}\,\propto\sigma^{+0.40}M_*^{\,0.00}$ & ${\rm \alpha/Fe}\,\propto\sigma^{+0.20}M_*^{\,0.00}$ 
& no $M_*$ dependence \\
\hline
Recovered $\sigma,L_R$ scalings 
&  ${\rm Z/H}\,\propto\sigma^{+0.41}L_R^{-0.05}$ & ${\rm Age}\,\propto\sigma^{+0.78}L_R^{-0.31}$ & ${\rm \alpha/Fe}\,\propto\sigma^{+0.20}L_R^{-0.00}$ & uncorrelated age--Z residuals & (1) \\
&  ${\rm Z/H}\,\propto\sigma^{+0.09}L_R^{+0.21}$ & ${\rm Age}\,\propto\sigma^{+0.71}L_R^{-0.25}$ & ${\rm \alpha/Fe}\,\propto\sigma^{+0.34}L_R^{-0.11}$ & perfect anti-correlation & (3) \\
&  ${\rm Z/H}\,\propto\sigma^{+0.17}L_R^{+0.14}$ & ${\rm Age}\,\propto\sigma^{+0.74}L_R^{-0.28}$ & ${\rm \alpha/Fe}\,\propto\sigma^{+0.33}L_R^{-0.10}$ & moderate anti-correlation & (2) \\
\hline
Observed  $\sigma,L_R$ scalings &  ${\rm Z/H}\,\propto\sigma^{+0.15}L_R^{+0.14}$ & ${\rm Age}\,\propto\sigma^{+0.70}L_R^{-0.24}$ & ${\rm \alpha/Fe}\,\propto\sigma^{+0.34}L_R^{-0.12}$ & (default aperture correction)\\
\hline
\end{tabular}
\end{table*}

\begin{figure*}
\includegraphics[angle=270,width=180mm]{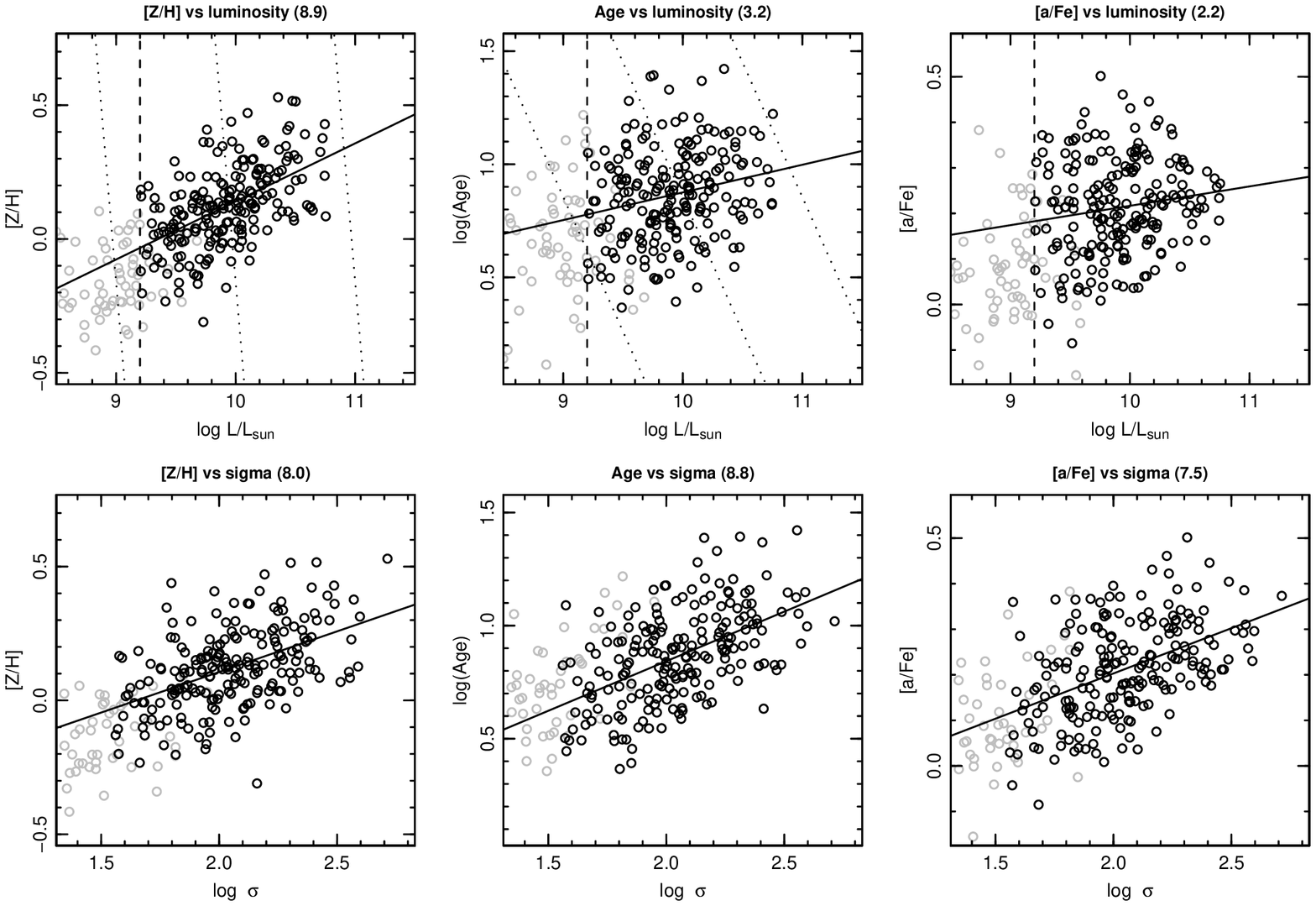} 
\caption{Stellar parameters versus two alternative mass proxies, as derived from our Monte Carlo model in which there is no intrinsic dependence on stellar mass. 
The figure shows a single realisation for $\sim$200 galaxies selected from the model to match the observed sample definition. The grey points indicate galaxies from the 
model that were excluded by the luminosity and $\sigma$ limits. The overall pattern of results should be directly compared with the observed data, in Figure~\ref{fig:fullcorplots}. 
}\label{fig:fullcorplots_sim}
\end{figure*}

We noted in Section~\ref{sec:proxychoice} that simple fits for correlations with $\sigma$ are compromised for samples selected by luminosity, 
and that although introducing luminosity as a second ``predictor'' leads to unbiased fits, these are difficult
to interpret because luminosity itself depends on the stellar populations. 
A physically more meaningful quantity to use as a predictor of age, etc, is the stellar mass, $M_*$. In this section, we
develop a Monte Carlo modelling approach to constrain the relative importance of $M_*$ and $\sigma$ in driving 
the stellar population scaling relations. 

\subsection{Modelling the scaling relations}

In principle, the stellar mass-to-light ratio $M_*/L$ can be derived from the spectroscopic age and metallicity 
estimates\footnote{For samples of galaxies having a wide intrinsic range in $M_*/L$, optical colour and near-infrared 
luminosity is widely used to estimate stellar mass (e.g. Bell et al. 2003). This method successfully discriminates between 
low-mass star-forming galaxies and more massive but passive objects. By contrast for old SSP models, i.e. {\it within} the passive galaxy population, 
optical colour is a very poor predictor of $M_*/L_K$, and hence this approach does not provide an improved estimate of $M_*$.}. 
However, it is not 
straightforward 
to use stellar masses derived this way to fit the scaling relations, because the
ages and stellar masses will have near-perfectly correlated measurement errors. A more general obstacle is that at low
stellar mass, the sample is necessarily biased by the luminosity selection towards younger and brighter galaxies. 

To estimate the correlations of age and metallicity with stellar mass, we instead resort to {\it modelling}
the assumed underlying distribution of galaxy properties such that a simulated luminosity-selected sample approximately reproduces
the data. In practice, this modelling requires {\it a priori} unknowns such as the distribution function of stellar 
masses and the dependence of velocity dispersion on stellar mass. Thus the models must be tuned to reproduce both
the luminosity distribution and the observed Faber--Jackson relation of the galaxy sample, as well as the observed
scalings of age and metallicity with luminosity and velocity dispersion. 

To construct a simulated galaxy sample, we first draw stellar masses from 
a Schechter function, with characteristic mass $M_*^\star=10^{10.7}M_\odot$, and low-mass slope $\alpha_{\rm MF}=-0.5$, and assign velocity dispersions with mean value following
$\log\sigma= 1.85 + 0.50 \log \frac{M_*}{10^{10}M_\odot}$, and a gaussian scatter of 0.125 in $\log\sigma$ around the mean. The numeric coefficients
here are chosen (but not rigorously fitted) to match the luminosity distribution and the observed Faber--Jackson relation of our galaxy sample. The adopted stellar mass function is in fact
similar to that derived for early-type (concentration-selected) galaxies by Bell et al. (2003), where $M_*^\star=10^{10.6}M_\odot$ and $\alpha_{\rm MF}=-0.6$. 
We attempt to reproduce the observed stellar population scaling relations 
$P=P_0 \ \sigma^a L^b$ in terms of underlying models of the form $P=P_0^\prime \ \sigma^{a^\prime} M_*^{b^\prime}$, where the $P$ are age, Z/H and $\alpha$/Fe. 
For given choices of model parameters 
$(P_0^\prime,a_P^\prime,b_P^\prime)$, we assign ages and metallicities, with a fixed scatter $s_P$ around the 
model. We allow the deviations from the mean scaling relation for age to be anti-correlated with those from the Z/H scaling relation, 
as proposed by Trager et al. (2000). The $\alpha$/Fe deviation is assumed to be correlated with the age deviation. We consider cases
of uncorrelated, moderately correlated and perfectly correlated deviates.
Using the Maraston (2005) SSP models, we next determine the stellar mass-to-light ratio for each simulated galaxy, based on 
its age and Z/H, and from this compute the luminosity $L_R$. (Abundance ratio effects on the mass-to-light ratio are neglected here.)
Finally, we impose selection criteria, $\log(L/L_\odot)>9.2$ and $\log\sigma>1.55$ to reproduce the
approximate limits of the observed galaxy sample. 

The $P=P_0 \ \sigma^a L^b$ relations can now be fit for the selected simulated galaxies, and compared to the observed correlations. 
Ideally, the model would be optimized to obtain confidence intervals for the parameters which
describe the underlying correlations (i.e. $P_0^\prime,a_P^\prime,b_P^\prime,s_P$, for each parameter $P$), marginalizing over
the ``nuisance parameters'' (i.e. those describing the stellar mass distribution function, the $M_*-\sigma$ relation, and the correlation structure
of the deviations from the mean scalings). In practice however, the high dimensionality precludes a full exploration of the solutions, at least within
this paper. 
Here, we simply test whether suitable solutions exist for underlying relations having no intrinsic dependence on stellar mass, 
i.e. all stellar population parameters are determined only by their velocity dispersion. Table~\ref{tab:massfits} and Figure~\ref{fig:lumsigfit2} summarize
the results obtained for input correlations of ${\rm Z/H}\propto\sigma^{+0.35}, {\rm Age}\propto\sigma^{+0.40}$, and $\alpha/{\rm Fe}\propto\sigma^{+0.20}$. 
As we will show, these input slopes are able to reproduce all of the observed trends with $\sigma$ and $L$, without 
invoking intrinsic stellar mass dependence. The required slopes were determined using a three-parameter grid search, with the parameters of the 
stellar mass distribution and the $M_*-\sigma$ relation held fixed, and adopting the case with ``moderate'' anti-correlation of age and metallicity residuals (see below). 
The scatter in parameters around the input model is constrained in all cases to reproduce the age, metallicity and \nafe\ scatter around the observed scaling
relations.

\subsection{Results}

In the case where the age and metallicity deviates are uncorrelated (solution 1 in Figure~\ref{fig:lumsigfit2} and Table~\ref{tab:massfits}), we find that although age does not depend explicitly on stellar mass, the recovered 
fit has age increasing very steeply with $\sigma$ but also decreasing with luminosity. This is primarily an effect of the variation in $M_*/L$: younger
galaxies are brighter. For Z/H and \nafe\, the recovered scaling relations are much closer to the input relations, because $M_*/L$ depends only weakly on
Z/H and not at all (in our modelling) on \nafe. 
Thus, if we do not allow for correlated deviations, our models with no intrinsic stellar mass dependence cannot match the observed trends in these parameters\footnote{
We estimate that to match the Z/H trends in the absence of correlated deviation, we would require approximately Z/H$\propto \sigma^{+0.1}M_*^{+0.2}$.
}.
Indeed, the recovered Z/H scaling would yield a luminosity slope opposite to that observed, because higher-metallicity galaxies have lower luminosities, other factors being equal.

Next we impose an anti-correlation between the age and metallicity deviations at fixed mass, such that galaxies that are younger than average are also more metal 
rich than average. 
This is consistent with the ``Z-plane'' in our own data (Smith et al. 2008) and with the conclusions reached in other work (e.g. Trager et al. 2000; Smith et al. 2009a). 
We also allow a positive correlation of the \nafe\ and age deviates. In the first instance we assume perfect anti-correlation among the residuals. In this case, the age trends are only
slightly altered from the uncorrelated-deviation solution. However, the recovered metallicity scalings are dramatically different: instead of the input pure $\sigma$ dependence, 
the observed scaling is almost purely with luminosity, in the sense that more luminous galaxies have higher Z/H. The explanation is that more luminous galaxies at given $\sigma$ and $M_*$
are younger, and the Z-plane now implies that they are also more metal rich. This indirect brightening effect is much larger than the direct dimming effect of higher Z/H. 
A similar effect holds for \nafe. The recovered scaling relations are now in much better agreement with the observed correlations, although for Z/H they ``overshoot'', producing slightly
too strong a luminosity trend. Adjusting the input $\sigma$ scaling does not much improve the match. 

Finally, noting that the observed trends lie intermediate between the results for uncorrelated and perfectly anti-correlated age--metallicity deviates, we assign stellar populations
to the simulated galaxies using moderately anti-correlated scatter ($\rho=-0.75$). With this prescription, we are able to match the observed scaling relations for all three 
stellar population parameters. 
Figure~\ref{fig:fullcorplots_sim} shows the simulated one-parameter scaling relations (i.e. equivalent to Figure~\ref{fig:fullcorplots}), for a realisation of our model, 
with moderate anti-correlation of the age/metallicity deviates. Exactly as in the observed galaxy sample, we recover metallicity scaling tightly both with luminosity and with $\sigma$, 
while age and \nafe\ follow $\sigma$ much more closely than luminosity. 
We emphasise that these predictions are derived from an underlying model in which stellar populations depend only on $\sigma$, 
and are independent of stellar mass. 
This result is robust to changes in the strength of the assumed metallicity gradients: e.g. if stronger aperture corrections are preferred, 
the degree of residual anti-correlation required is reduced. 
Note also that the scalings with velocity dispersion in the Monte Carlo models (e.g. Age\,$\propto\sigma^{+0.40}$) are not substantially different
from the single-parameter fits of Section~\ref{sec:lumsigscals} (where Age\,$\propto\sigma^{+0.44\pm0.05}$). 
In principle, the modelling here explicitly takes into account the biases described in Section~\ref{sec:proxychoice}. The similarity to the
one-parameter scalings confirms that selection biases do not strongly influence the simpler fitting method (as found also by Allanson et al. 2009). 

\begin{figure}
\includegraphics[angle=270,width=80mm]{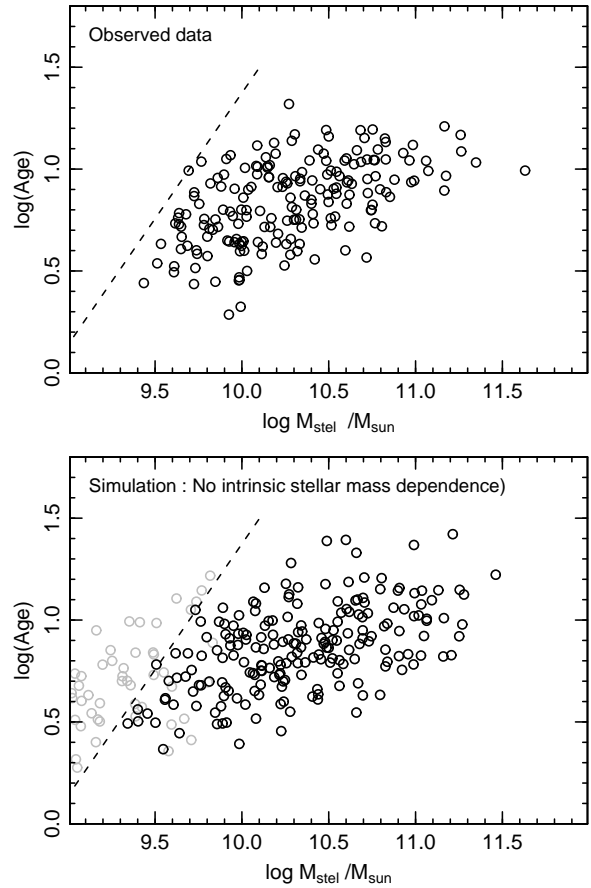} 
\caption{The age versus stellar mass relation in the observed sample (above) compared to a representative realisation from 
the simulation (below), in which age$\,\propto\sigma^{+0.40}$ but there is no intrinsic dependence on $M_*$. 
For the observations, stellar masses have been estimated using the observed ages and metallicities. The dashed line in each panel indicates
the $10^{9.5}\,M_\odot$ luminosity limit, assuming a fixed metallicity. 
For the simulated data, the figure assumes realistic measurement errors in the age and metallicity. The stellar masses shown are
those which would be computed from the spectra by an observer, and hence include the effects of these measurement errors. 
Grey points indicate simulated galaxies excluded by the luminosity and $\sigma$ limits.}
\label{fig:rev_masstest}
\end{figure}

With the simulations in hand, we can assess the apparently simpler approach of fitting scaling relations using stellar masses estimated
directly from the measured ages and metallicities. As noted above, this is complicated by the very strong correlation between age and
spectroscopic stellar mass, and by the luminosity selection limit, which translates into a sharp diagonal cut in the age--$M_*$ plane. 
Figure~\ref{fig:rev_masstest} shows the age--$M_*$ relation for the observed sample, and for a realisation of our simulation, in which there is no
explicit dependence on stellar mass. In both cases a strong correlation is observed, but this primarily reflects the correlation with velocity dispersion, 
and bias from the luminosity limit. 
Fitting simultaneously using $\sigma$ and $M_*$, the observed data yield Age\,$\propto\sigma^{+0.26\pm0.10}M_*^{+0.13\pm0.05}$. Hence naively we would 
infer a significant correlation of age with stellar mass, at fixed velocity dispersion. However, a similar correlation is recovered for our 
simulated samples, which have no intrinsic $M_*$ dependence. Specifically, we obtain Age\,$\propto\sigma^{+0.24}M_*^{+0.09}$ which is indistinguishable 
from the measured trends. In both the simulated and the observed samples, the apparent dependence on $M_*$ (at fixed velocity dispersion) arises from
error correlations and from the luminosity selection bias. 

We conclude that for reasonable assumptions about the galaxy population (mass distribution, $M_*-\sigma$ relation,  Z-plane anti-correlation of age and metallicity residuals at fixed $\sigma$), 
we can reproduce the observed scaling relations from a model in which stellar populations depend only on velocity dispersion, and not on stellar mass.  
A residual correlation with $M_*$ can not be {\it excluded}, since the assumptions entering into the model are uncertain to an extent, and a full exploration of the 
parameter space has not yet been performed. Rather, our conclusion is that the data do not at present {\it require} any stellar mass dependence, 
in contrast to recent claims, which we discuss in the following section.

\section{Discussion}\label{sec:discuss}

During preparation of this paper, Graves et al. (2009, hereafter GFS09) published an analysis of composite spectra from the SDSS which 
addresses some of the issues that are discussed here. At an observational level, the results of GFS09 agree closely with the fits we obtain
from our sample of galaxies in Shapley. However, our interpretation of the results, and especially our conclusion that no residual dependence
on $M_*$ is required, differ from that of GFS09, as we discuss in this section. 

GFS09 constructed composite spectra for 16\,000 SDSS galaxies with $0.04<z<0.08$, binned by $\sigma$, $L$ and optical colour. 
The high-S/N stacked spectra for each bin were used to derive SSP-equivalent parameters, via comparison to models from Schiavon (2007), 
which were then used to explore how the stellar populations are correlated with luminosity and velocity dispersion. 
Their galaxy sample was selected to have concentrated $r^{1/4}$-like luminosity profiles, as well as to have no detected emission lines, contrasting with
the absence of explicit morphological cut in our sample. 
The total range of galaxy mass is narrower in GFS09 than here, with $\log\sigma>1.86$ and $\log(L/L_\odot)>9.4$. Moreover the
sample is drawn from the general galaxy population, rather than being explicitly a cluster-galaxy sample as here. 
GFS09 did not apply any correction for metallicity gradients, stating that the SDSS 3-arcsec fibres are ``not nuclear spectra but sample
a substantial fraction of the galaxy light''. 

Using the stellar population parameters for the stacked spectra, GFS09 recover positive correlations of age, Fe/H and Mg/Fe both with  velocity dispersion and with luminosity. 
In agreement with our results, the correlations for age and Mg/Fe are much stronger when $\sigma$ is the predictor variable, while Fe/H is more strongly
correlated with luminosity.  GFS09 did not provide quantitative descriptions of these correlations. Fitting to the data in their tables 1 and 2, we obtain 
Age$\,\propto$\,$\sigma^{+0.95}L^{-0.25}$, 
Fe/H$\,\propto$\,$\sigma^{-0.10}L^{+0.28}$ and
Mg/Fe$\,\propto$\,$\sigma^{+0.53}L^{-0.12}$. 
For a fair comparison to our results for metallicity, we use the translation derived by Smith et al. (2009a) to convert from Fe/H and Mg/Fe in the Schiavon models
to Z/H in the TMBK models. Fitting the resulting quantity yields Z/H$\,\propto\sigma^{+0.30}L^{+0.21}$. These results are qualitatively similar to ours, although outside our formal error 
ellipses mainly due to a stronger $\sigma$ dependence for Mg/Fe, which affects both the \nafe\ and Z/H comparisons. 

Although we broadly agree with GFS09 in terms of the {\it observed} correlations, we differ in our {\it interpretation} of the luminosity component of the trends. 
GFS09 claimed that the variations in $M_*/L$ due to varying age are not sufficient to generate the luminosity effects observed, and asserted that the luminosity
dependence of the stellar populations reflects a real correlation with stellar mass, rather than ``trivial'' effects due to $M_*/L$. 
In Section~\ref{sec:stelmass}, we explicitly examined the effects of $M_*/L$ variations through Monte Carlo simulations including the effects of sample selection, as well as
intrinsic correlations between age and metallicity at fixed mass. We showed that, for our observational strategy, stellar populations which depend only on $\sigma$ 
can give rise to luminosity dependence consistent with that observed. It seems plausible that the similar trends found by GFS09 can be explained similarly, but
to confirm this would require equivalent Monte Carlo modelling tuned to the different strategy used in their work. 

Moreover, we have seen that the 
effect of metallicity gradients can be substantial when deriving trends with $\sigma$ and $L$ simultaneously, because the Fundamental Plane implies that the brightest
galaxies at a given velocity dispersion have larger effective radius, and are thus subject to larger aperture corrections. 
The {\it range} of aperture corrections at fixed $\sigma$, over a given interval in luminosity, is independent of fibre size and independent of distance to the galaxy.
Thus despite GFS09 using larger fibres, and covering a significant range in distance, we expect that uncorrected metallicity gradients will generate a bias in
their Fe/H scaling relations, in favour of a steeper slope with $L$, and a shallower dependence on $\sigma$.

Finally, we note that our results partly disagree with those of Gallazzi et al. (2006), who found that the anti-correlation of age and luminosity at fixed $\sigma$ in
an SDSS-based sample could not be entirely accounted for by variations in $M_*/L$ and hence some residual dependence with $M_*$ was required. For metallicity, however, Gallazzi et al. 
conclude that dynamical mass is more important than stellar mass, in accordance with our results.

\section{Conclusions}\label{sec:concs}

In this paper we have used previously-reported absorption line strengths for passive galaxies in the Shapley Supercluster
to obtain estimates of SSP-equivalent age, metallicity and $\alpha$-element abundance ratios. Typical formal errors on the derived
ages and metallicities are $\sim10$\,per cent. The derived population parameters reproduce the observed broadband colours, with a weak 
residual colour dependence on \nafe\ in the sense predicted by stellar evolution models. Using the Rose \caii\ index, 
we have shown that the young  (2--5\,Gyr) SSP-equivalent ages in some galaxies do not, in general, result from a low mass-fraction ``frosting'' 
of star-formation within the past Gyr. 

We investigated the variation of stellar population parameters as a function of both velocity dispersion and luminosity.
Examining both the residual correlations after fitting one variable, and the bivariate fits, we find strikingly different behaviours for
the different parameters. Age and \nafe\ are correlated primarily with $\sigma$, but with an additional trend with luminosity such that
at given $\sigma$, the more luminous galaxies are younger and less $\alpha$-enriched. For total metallicity Z/H, the observed correlations
appear to be primarily with luminosity. These observed trends, for galaxies in rich cluster cores, are in close agreement with the recent 
results of Graves et al. (2009) for a general sample of galaxies from SDSS. Thus it is unlikely that the results are strongly dependent on
the environments that are sampled. 

The luminosity trends would naively suggest that metallicity, in particular, may be primarily correlated with stellar mass, rather than 
with the depth of the potential well. Such a result appears contrary to galactic wind models for the mass-metallicity relation. 
However, using Monte Carlo simulations, we showed that the apparent luminosity correlations can be recovered from an underlying model
in which stellar populations depend only on velocity dispersion, and not at all on stellar mass. 
Thus the observed luminosity trends 
{\it do not imply corresponding trends of stellar population parameters with stellar mass}. 
For age, the luminosity correlation arises because 
young galaxies are brighter, at fixed $M_*$. For metallicity, the observed luminosity 
dependence arises because metal-rich galaxies, at fixed mass, tend also to be younger, and hence brighter. 
The presence of metallicity gradients also boosts the apparent luminosity dependence, if aperture corrections are not applied. 

After accounting for these effects, we conclude that the stellar populations of passive galaxies 
are fundamentally correlated with velocity dispersion, and hence with the depth of their gravitational potential well.
There is no convincing evidence for additional dependence on the stellar mass, after accounting for the velocity dispersion correlations.

\section*{Acknowlegments}

RJS was supported for this research under the STFC rolling grant PP/C501568/1 `Extragalactic Astronomy and Cosmology 
at Durham 2005--2010'. We thank Chris Haines and the SOS team for providing the $B-R$ colours used in 
Section~\ref{sec:coltest}. We are grateful to the anonymous referee for providing a constructive and thoughtful report
on the manuscript. 

{}

\label{lastpage}

\end{document}